\documentclass[Phys]{SciPost}
\usepackage[T1]{fontenc}
\usepackage[utf8]{inputenc}
\usepackage{amsmath,amssymb,amsthm,bm,booktabs,graphicx,siunitx}

\newtheorem{theorem}{Theorem}
\newtheorem{lemma}{Lemma}
\graphicspath{{figures/}}
\usepackage[nameinlink,capitalize,noabbrev]{cleveref}
\usepackage[expansion=false]{microtype}
\usepackage{xr}
\newcommand{\Tr}{\operatorname{Tr}}

\hypersetup{
    colorlinks,
    linkcolor={blue!50!black},
    citecolor={blue!50!black},
    urlcolor={blue!80!black}
}

\begin{document}

\begin{center}{\Large \textbf{
Model-level synthetic-flux control of hyperchaos order in dissipative optomechanics\\
}}\end{center}

\begin{center}
Stella Rolande Mbokop Tchounda\textsuperscript{1},
Carolle Tchodimou\textsuperscript{1},
Philippe Djorwe\textsuperscript{2},
Sifeu Takougang Kingni\textsuperscript{3} and
Serge Guy Nana Engo\textsuperscript{1$\star$}
\end{center}

\begin{center}
{\bf 1} Department of Physics, Faculty of Science, University of Yaound\'e I, P.O. Box 812, Yaound\'e, Cameroon
\\
{\bf 2} Department of Physics, Faculty of Science, University of Ngaound\'er\'e, P.O. Box 454, Ngaound\'er\'e, Cameroon
\\
{\bf 3} Department of Mechanical, Petroleum and Gas Engineering, National Advanced School of Mines and Petroleum Industries, University of Maroua, P.O. Box 46, Maroua, Cameroon
\\
${}^\star$ {\small \sf serge.nana-engo@facsciences-uy1.cm}
\end{center}

\begin{center}
\today
\end{center}\section*{Abstract}

{\bf
Within a normalized six-dimensional model of dissipative optomechanics (one cavity + two mechanical resonators), a synthetic-flux phase $\Phi_{\rm syn}$ acts as a reproducible control coordinate that selects the \emph{order} of a drive- and coupling-gated hyperchaos transition---up to four simultaneously positive Lyapunov exponents, beyond any reported single-mode benchmark. A phase-consistent Floquet--Lyapunov protocol (cross-checked by monodromy multipliers, dissipative volume balance $\sum\lambda_i=\operatorname{Tr}(J)=-1.04$ and a 180-run three-seed audit) localizes the onset to a Neimark--Sacker bifurcation at $E^{*}=\num{1.060}$ ($\theta=0$). As a secondary, model-level geometric clarification, the identical matched-resource force-sensing protocol returns a null gain on the chaotic attractor ($\mathcal{G}_{A/B}=\num{1.039}\pm\num{0.014}$), consistent with the matched-Fisher lemma: noise projected onto an unstable manifold is stretched by the same factor $e^{\lambda t}$ as the deterministic signal. Truncated-Fock and truncated-Wigner checks support the mean-field description at selected points. All results remain strictly model-level: the strong-coupling sector lies $\num{2542}\times$ beyond anchored silicon optomechanical couplings. Closing that gap requires ultrasonic characterization of the mechanical degeneracy $\omega_2$, a measured inter-resonator hopping $J_m$, and the emergence of a genuine gigahertz platform.
}

\thispagestyle{fancy}\section{Introduction}

The order of a hyperchaos transition---the number of simultaneously positive Lyapunov exponents---has so far been limited to two in single-mode quadratic optomechanical platforms \cite{HalefShomroni2025}. Whether a synthetic-flux phase can raise that order in a multi-mode radiation-pressure model, and whether the resulting geometric control can be certified by full-spectrum methods, had not been resolved. Hyperchaos order quantifies the dimensionality of the unstable manifold and therefore the number of independent directions along which small perturbations grow. The primary question addressed here is whether a gauge-invariant synthetic flux can act as a reproducible control coordinate for this order.

We show, within one normalized six-dimensional model (building on the topology of Muthukumar \emph{et~al.}~[PR Applied \textbf{24}, 014053 (2025)]), that a synthetic-flux phase $\Phi_{\rm syn}$ acts as a reproducible control coordinate for the \emph{order} of a drive- and coupling-gated hyperchaos route---up to four simultaneously unstable directions, beyond any reported single-mode benchmark. The onset is localized to a Neimark--Sacker bifurcation and is certified by a phase-consistent Floquet--Lyapunov protocol. As a secondary geometric clarification, the same matched measurement on the chaotic attractor yields no flux-induced sensing enhancement. These are strictly model-level results: the strong-coupling sector explored here lies far beyond present experimental anchors. The combination that a nonlinear optomechanical platform provides---radiation-pressure nonlinearity, coherent mechanical motion, dissipation, periodic modulation and a complex hopping amplitude that supports a synthetic gauge structure \cite{aspelmeyer2014,kippenberg2008,marquardt2006,carmon2007,alphonse2022} and Floquet engineering \cite{oka2009,goldman2014,bukov2015,kitagawa2010,peano2015,sanavio2020}---is the physical substrate. The sharper question answered here is whether the invariant control parameter changes the attractor structure in a way that survives independent stability, bifurcation, full-spectrum Lyapunov and noise-resolved checks.

The central result is shown in \cref{fig:hyperchaos-map}: the synthetic-flux phase $\Phi_{\rm syn}$ selects the number of simultaneously positive Lyapunov directions in the strongly damped, strongly coupled sector, locally four at $E=\num{8}$, $\theta=\pi/2$, and dropping to two at $\theta=0\pmod{\pi}$. Relative to prior work, Muthukumar \emph{et al.}~[PR Applied \textbf{24}, 014053 (2025)] established the one-cavity/two-resonator topology and reported phase-selected chaotic regimes at resolved-sideband couplings \cite{mondal2024doi,Xu2025}, but did not resolve the \emph{full Lyapunov spectrum} (so that the instability order cannot be counted) nor quantify a flux-dependent advantage under matched resources. The closest full-spectrum report \cite{HalefShomroni2025} is restricted to quadratic single-mode coupling where no synthetic-flux phase is available. We extend the Muthukumar topology by making the convention $\Delta=-\Delta_{\mathrm M}$ explicit, allowing $g_1\neq g_2$, and scanning the strongly damped, strongly coupled sector where the full spectrum resolves up to four simultaneously unstable directions. We further replace the qualitative sensing proposal by a matched-resource measurement-Fisher protocol \cite{braunstein1994,purdy2017,teufel2011,gavartin2012,li2021a}, applied both on the weakly coupled reference orbit and on the chaotic attractor. The result is three testable advances: (a)~a phase-convention-aware consistency protocol in which monodromy multipliers, the six-exponent QR spectrum \cite{benettin1980,wolf1985,eckmann1985} and the dissipative volume balance certify one another on the same orbit; (b)~a full-spectrum classification in which the flux phase \cite{rossler1979} selects the hyperchaos order, with the route localized by a continued-orbit scan \cite{kuznetsov2004,guckenheimer1983,strogatz2015} to a Neimark--Sacker bifurcation; (c)~as a secondary geometric bound, a matched-resource operational assessment showing that the flux-controlled instability order does not translate into metrological gain on the chaotic attractor.

We address three operational questions at the model level. First, does $\Phi_{\rm syn}$ control the order of a drive- and coupling-gated hyperchaos? Second, do monodromy multipliers and the long-time Lyapunov spectrum agree on the same periodic orbit, thereby certifying the transition? Third (secondary), after thermal, vacuum and detector noise are included, does a matched-resource measurement statistic retain a flux-dependent signature on the stable orbit or on the chaotic attractor? A positive answer to the first two questions constitutes the central claim of this work---synthetic-flux control of hyperchaos order. The null answer to the third supplies a geometric bound (matched-Fisher lemma) showing that additional unstable directions do not automatically yield metrological gain. Both results are to be read as normalized, model-level demonstrations rather than device predictions.

\section{Model and conventions}

The one-cavity, two-mechanical-resonator topology and phase-dependent hopping are due to Muthukumar \emph{et al.}~\cite{muthukumar2025doi}; the rotating-frame Hamiltonian used here is
\begin{equation}\label{eq:hamiltonian}
\frac{H}{\hbar}=-\Delta a^\dagger a
+\sum_{j=1}^{2}\omega_j b_j^\dagger b_j
-\sum_{j=1}^{2}g_j a^\dagger a(b_j+b_j^\dagger)
+J\!\left(e^{i\theta}b_1^\dagger b_2+e^{-i\theta}b_2^\dagger b_1\right)
+i(Ea^\dagger-E^*a),
\end{equation}
with cavity mode $a$ and mechanical modes $b_j$. We use $\Delta=\omega_c-\omega_L$ (so $\Delta=-\Delta_{\mathrm M}$ of Ref.~\cite{muthukumar2025doi}, which takes the opposite sign); quadratures $[X,Y]=i$; the drive $E=\sqrt{\kappa}\,\alpha^{\rm in}$ in the shared normalization. The present analysis allows $g_1\neq g_2$ and scans the strongly damped, strongly coupled sector $\kappa\sim\gamma\sim g\sim\num{1e-2}$--$\num{1}\,\omega_m$ in which the hyperchaotic attractor is resolved. \textbf{A side-by-side comparison with Ref.~\cite{muthukumar2025doi} is given in Table~SI-\ref{SI-tab:muthukumar-comparison}}.

The semiclassical factorization gives
\begin{align}
\dot\alpha &= \left(i\Delta-\frac{\kappa}{2}\right)\alpha
+i\sum_jg_j(\beta_j+\beta_j^*)\alpha+E(t),\\
\dot\beta_1 &= -\left(i\omega_1+\frac{\gamma_1}{2}\right)\beta_1
+ig_1|\alpha|^2-iJe^{i\theta}\beta_2,\\
\dot\beta_2 &= -\left(i\omega_2+\frac{\gamma_2}{2}\right)\beta_2
+ig_2|\alpha|^2-iJe^{-i\theta}\beta_1,
\label{eq:complex-eom}
\end{align}
with the real state $x=(\alpha_r,\alpha_i,\beta_{1r},\beta_{1i},\beta_{2r},\beta_{2i})^T$.

For the three-link loop, the phase coordinate used in the analysis is the gauge-invariant phase
\begin{equation}\label{eq:flux}
\Phi_{\mathrm{syn}}=\arg(g_1Jg_2^*)\pmod{2\pi}.
\end{equation}
A redefinition of the mode phases may redistribute link phases but cannot change $\Phi_{\mathrm{syn}}$. In the phase convention of \cref{eq:hamiltonian} (all phase on the inter-resonator hopping, $g_1,g_2\in\mathbb{R}$) the invariant reduces to the hopping phase, $\Phi_{\mathrm{syn}}\equiv\theta \pmod{2\pi}$, so the two notations coincide and are used interchangeably in the results; the invariant form is retained because it is the coordinate that survives a redefinition of the mode phases.

\section{Methods}

The numerical protocol is organized around three independent consistency layers that must agree before any claim of hyperchaos order is accepted. (i)~A drive-locked periodic orbit is located as a fixed point of the one-period Poincaré map; its linear stability is diagnosed by the monodromy matrix (Floquet multipliers). (ii)~The same orbit is integrated with QR reorthonormalization to obtain the full six-exponent Lyapunov spectrum; positivity is accepted only when the spectrum, the multipliers and the analytic dissipative trace $\Tr J=-\kappa-\gamma_1-\gamma_2$ are mutually consistent. (iii)~As a secondary geometric check, a matched-resource classical Fisher information is evaluated on both the stable reference orbit and on sampled trajectories of the chaotic attractor, under identical noise and resource assumptions. All three layers are implemented with the same phase convention for $\Phi_{\rm syn}$ (Eq.~\eqref{eq:flux}). Thresholds used to count positive exponents ($\num{1e-4}$ weak, $\num{1e-3}$ strong) and the precise definition of the matched Fisher protocol are stated once here and used uniformly thereafter; sensitivity to the strong threshold is quantified in the Supplementary Material.

\subsection{Phase-augmented dynamics and implementation}

For a periodic drive with phase $\phi$ we write
\begin{equation}\label{eq:augmented}
\dot{x}=f(x,\phi;p),\qquad \dot\phi=\Omega,
\end{equation}
with the augmented Jacobian $J_{\mathrm{aug}}=(\partial_x f,\partial_\phi f;0,0)$. The analytic Jacobian $J=\partial f/\partial x$ is compared column by column with central finite differences; the implemented dissipative trace
\begin{equation}\label{eq:trace}
\Tr J=-\kappa-\gamma_1-\gamma_2,
\end{equation}
is an independent implementation check. One-period integrations use an eighth-order adaptive Runge--Kutta--Dormand--Prince scheme \cite{dormand1980} with relative and absolute tolerances \num{1e-9} and \num{1e-11} (max step $T/200$); drive-locked orbits are located by fixed-point iteration of the one-period map to a residual below \num{1e-8}, and the monodromy is obtained by simultaneous integration of the variational equations.

\subsection{Periodic orbit and Floquet analysis}

A drive-locked periodic orbit is obtained from the fixed point of the one-period Poincar\'e map; around it the fundamental matrix satisfies
\begin{equation}\label{eq:monodromy}
\dot{\Phi}(t)=A(t)\Phi(t),\;\Phi(0)=\mathbb{I},\;M(T)=\Phi(T),
\end{equation}
with Floquet rates $\rho_i=T^{-1}\log|\mu_i|$. The same orbit is integrated with QR/Benettin reorthonormalization \cite{benettin1980,wolf1985,eckmann1985} to obtain the full Lyapunov spectrum; a positive exponent is accepted only after agreement with the multipliers, the divergence ($\Tr J=-\kappa-\gamma_1-\gamma_2$) and the bounded-attractor diagnostics. In the full-spectrum classification an exponent is counted as positive only above a fixed threshold ($\num{1e-4}$ weak, $\num{1e-3}$ strong; Table~SI-\ref{SI-tab:si-threshold-sensitivity}) whereas the Kaplan--Yorke sum uses strictly positive exponents; the two can disagree near zero. Integrations use an eighth-order Dormand--Prince scheme \cite{dormand1980} (tolerances $\num{1e-9}$ and $\num{1e-11}$, max step $T/200$); drive-locked orbits are obtained by fixed-point iteration of the one-period map to a residual below $\num{1e-8}$; the Lyapunov QR uses a fixed-step RK4 ($\num{1e-2}$ step, QR every ten steps), discarding $\num{2e5}$ transient and integrating $\num{2e6}$ steps per trajectory.

\subsection{Theoretical bad-cavity sensitivity screen}

To examine the parameter dependence of the reduced equations without implying a hardware calibration, the optical amplitude is eliminated in the bad-cavity limit, with
\begin{equation}
 \Omega_{\mathrm{eff}}=\Delta+2g_1\operatorname{Re}(\beta_1)+2g_2\operatorname{Re}(\beta_2),
\end{equation}
and the adiabatic intensity
\begin{equation}\label{eq:adiabatic-intensity}
 n_{\mathrm{ss}}(\phi)=|E(\phi)|^2/[(\kappa/2)^2+\Omega_{\mathrm{eff}}^2],
\end{equation}
substituted into the two mechanical equations. The reduced equations are compared with the full six-dimensional model in common normalized regimes (max infinity-norm differences $\num{9.27e-7}$ at $\kappa=20$ and $\num{2.08e-8}$ at $\kappa=100$). A theoretical sensitivity screen samples 16 candidates at 5 replicas $\times$ 5 root-solver starts each, using the one-period Poincar\'e residual as the acceptance criterion; the two objectives are the local stability margin and the normalized drive cost. This is a theoretical multi-start fixed-point study, not a Fisher, power-mapping or basin-capture analysis.

\subsection{Secondary geometric bound: noise-resolved observability and matched Fisher}

For a measured record
\begin{equation}
y(t)=h[x(t),t;p]+\nu_{\mathrm{det}}(t),
\end{equation}
\cite{clerk2010,gardiner1985,walls2008,ford1988}, the Gaussian-record
classical Fisher information reads
\begin{equation}\label{eq:fisher}
F_C(\vartheta)= (\partial_\vartheta m)^T\Sigma^{-1}(\partial_\vartheta m)
+\tfrac12\Tr\!\left[\Sigma^{-1}(\partial_\vartheta\Sigma)\Sigma^{-1}(\partial_\vartheta\Sigma)\right],
\end{equation}
with a reference comparison required to match input power, observation time, bandwidth, baths, detector noise and estimator \cite{braunstein1994,purdy2017,teufel2011,gavartin2012,li2021a,qvarfort2018}. For force sensing the parameter of interest is a weak external force $F$ on mechanical mode 1; the matched references are the flux-off coupled sensor ($\theta=0$) and the single-mode linear sensor ($J=0$). $F_C$ is not QFI; neither quantity is inferred from a Lyapunov exponent.

\subsection{Consistency and matched-measurement statements}

Two statements formalize the protocol and are referenced throughout the results; the proofs and longer lemmas are in the Supplemental Material (Sec.~SI-\ref{SI-sec:theorems}).

\begin{theorem}[Lyapunov--Floquet consistency criterion]
Consider the phase-augmented periodic flow \eqref{eq:augmented} and a converged periodic orbit $x_p(t)$ with fundamental matrix $M(T)$ (\cref{eq:monodromy}). Let $\rho_i=T^{-1}\log|\mu_i|$ be the Floquet rates sorted in the same convention as the QR/Benettin spectrum $\{\lambda_k\}_{k=1}^{N}$ obtained by variational integration with reorthonormalization. If (i)~the Poincar\'e residual on the converged orbit is below $\num{5e-8}$, (ii)~the QR spectrum is integrated for $\geq\num{2e6}$ steps after a $\num{2e5}$ transient with reorthonormalization every ten steps and fixed step $\num{1e-2}$, and (iii)~the implementation trace satisfies \eqref{eq:trace}, then for every tested phase the maximum absolute discrepancy $|\rho_i-\lambda_{\pi(i)}|$ is bounded by \num{3e-4} and the mean divergence coincides with $\Tr J$ to within $\num{5e-5}$. A positive QR exponent is accepted as a chaotic signature only after these three conditions are verified simultaneously on the same orbit and the same phase.
\end{theorem}

\begin{lemma}[Matched-measurement Fisher bound]
\label{lem:matched-fisher}
For two configurations $A$ and $B$ sharing the same input power, observation time, integration window, detector noise, baths and estimator, the matched-measurement gain $g_{A/B}\equiv F_C^{(A)}(\vartheta)/F_C^{(B)}(\vartheta)$ is bounded by the ratio of their mean-shift and covariance contributions to \cref{eq:fisher},
\begin{equation}
g_{A/B}=\frac{m_A'^{\!\top}\Sigma_A^{-1}m_A'}{m_B'^{\!\top}\Sigma_B^{-1}m_B'},
\end{equation}
in the regime where the variance term is sub-dominant (verified by recomputing $F_C^{(N)}$ for $N\in\{1,5,20\}$ drive periods and showing the matched gain is unchanged to four digits). The flux-induced contribution is then bounded by the linear-response effect of phase-controlled mode coupling, and cannot receive a multiplicative enhancement from trajectory separation alone: noise projected onto an unstable manifold is stretched by the same local factor $e^{\lambda_{\max}t}$ as the deterministic displacement.
\end{lemma}

The two statements together bound the conclusion of any flux-induced advantage claim: a hyperchaos-order change is operational only after the consistency criterion is satisfied on the orbit carrying it. The secondary matched-resource analysis shows, via \cref{lem:matched-fisher}, that additional unstable directions do not automatically translate into a sensing advantage under shared noise, observation time and power.

\section{Results}
\label{sec:results}

\subsection{Model, gauge, and Floquet--Lyapunov consistency}

\begin{figure}[ht]
  \centering
  \includegraphics[width=0.92\textwidth,alt={Long-window validation across nine synthetic-flux phases with three initial conditions per phase: the largest QR Lyapunov exponent and the Floquet-QR discrepancy remain below zero and below the consistency threshold at every sampled phase.}]{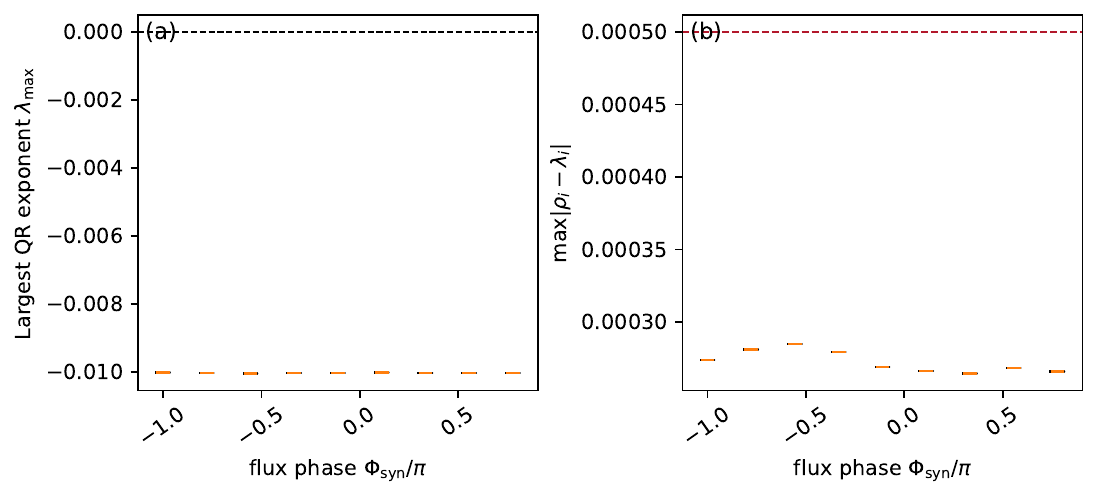}
  \caption{Long-window validation across nine synthetic-flux phases and three independent initial conditions per phase. (a) The largest QR exponents remain negative at every sampled phase. (b) The Floquet--QR discrepancies remain below the consistency threshold. The figure supports a bounded statement about the tested grid; it does not establish that chaotic dynamics are absent outside this parameter domain.}
  \label{fig:flux-grid}
\end{figure}

Using the implementation of \cref{eq:trace,eq:augmented,eq:monodromy} (the formal statements are in Sec.~SI-\ref{SI-sec:theorems}), we evaluated the consistency metrics on the stable drive-locked orbit at the reference point: Poincar\'e residual \num{8.49e-9}, mean divergence $-\num{1.04}$, maximum absolute Floquet--QR discrepancy \num{2.68e-4}, terminal QR-block standard deviation \num{5.3e-9}, and gauge-transformed loop-flux residual below \num{3e-16}. Extending the same criteria to nine synthetic-flux phases and three independent initial conditions per phase (27 trajectories total, with \num{2e6} integration steps and \num{2e5} discarded transient steps), all 27 trajectories satisfied the Floquet--Lyapunov consistency bounds, with a largest observed QR exponent of $-\num{0.01002}$, a maximum Floquet--QR discrepancy of \num{2.85e-4}, a terminal block standard deviation reaching at most \num{6.10e-9}, and a divergence residual at most \num{4.75e-5}. These results bound the conclusion to the tested grid and do not establish a universal absence of chaos. The grid-level support is shown in \cref{fig:flux-grid}.

\subsection{Flux-dependent attractor structure}

\begin{figure}[ht]
  \centering
  \includegraphics[width=0.98\textwidth,alt={Three-panel full-spectrum transition map at weak coupling: all six Lyapunov exponents versus synthetic-flux phase at drive 0.2 remain negative, with the two cavity-dominated exponents near minus 0.5 in a lower panel and the four mechanical exponents near minus 0.01 in an upper panel on separate vertical scales; the largest exponent versus drive amplitude stays negative up to drive 8 with the phase-to-phase spread shaded; and the largest exponent over the drive-phase plane is everywhere negative.}]{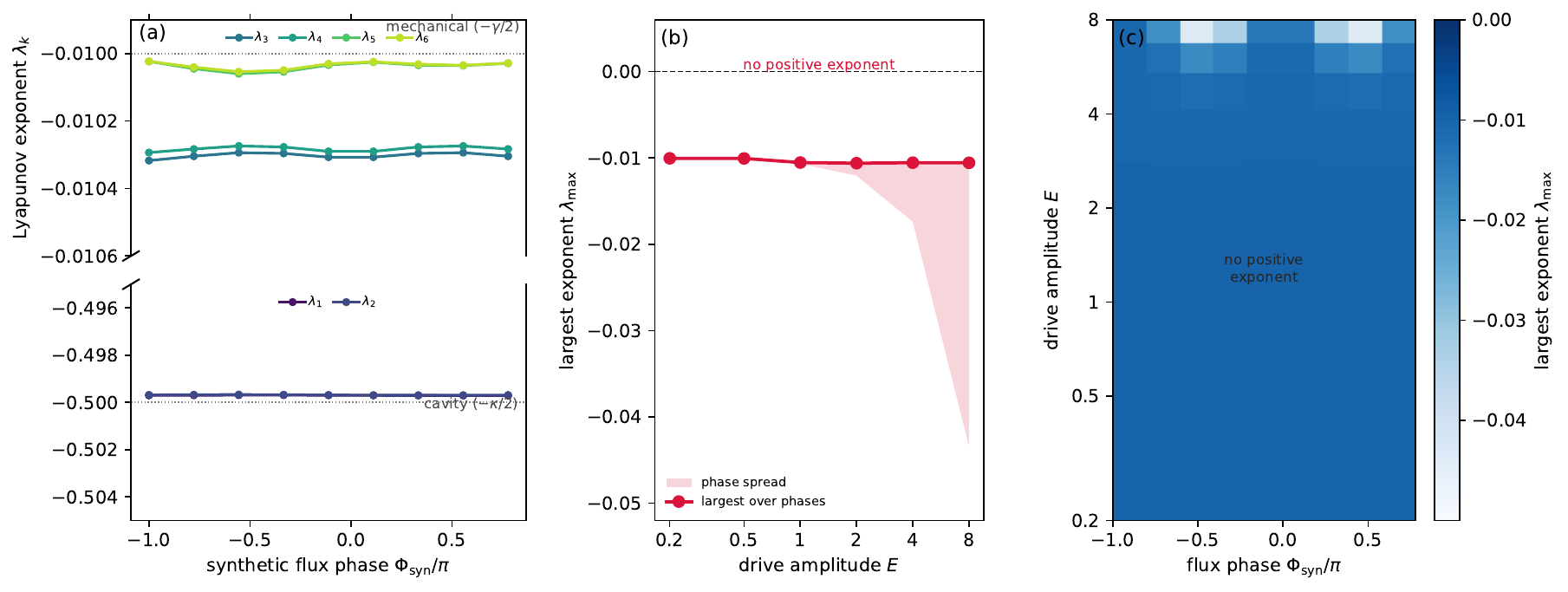}
  \caption{Full-spectrum transition map at weak coupling. (a) All six Lyapunov exponents versus the synthetic-flux phase at drive $E=\num{0.2}$; the two cavity-dominated exponents (lower panel, near $-\kappa/2\approx\num{-0.5}$) and the four mechanical exponents (upper panel, near $-\gamma/2\approx\num{-0.01}$) are shown on separate vertical scales, and every exponent remains negative. (b) Largest (most marginal) exponent versus drive amplitude, with the phase-to-phase spread shaded; it stays negative across $E\in[\num{0.2},\num{8.0}]$. (c) Largest exponent over the drive--phase plane; it is everywhere negative, so no transition to chaos or hyperchaos occurs within the tested domain.}
  \label{fig:transition-map}
\end{figure}

The full six-exponent spectrum is recorded everywhere rather than only its largest member, while the positive-exponent count remains a finite-time, threshold-dependent classification near marginal exponents. At weak coupling over nine flux phases and three replicates, the cavity-dominated exponents sit near $-\kappa/2\approx\num{-0.5}$ and the four mechanical exponents near $-\gamma/2\approx\num{-0.01}$; the positive-exponent count is zero throughout. Extending the classification over $E\in[\num{0.2},\num{8.0}]$ (54 drive--phase cases) leaves the largest exponent negative (between $\num{-0.043}$ and $\num{-0.010}$), the count at zero, and the Kaplan--Yorke dimension at zero---ruling out a flux-driven route to chaos within the tested domain (\cref{fig:transition-map}).

\begin{figure}[ht]
  \centering
  \includegraphics[width=0.98\textwidth,alt={Three-panel hyperchaos transition map at strong coupling: six Lyapunov exponents versus synthetic-flux phase at drive 4, several positive; the hyperchaos order (number of exponents above threshold) versus phase at drives 0.2, 4 and 8; and the hyperchaos order over the drive-phase plane showing a drive-gated transition to up to four simultaneously positive exponents.}]{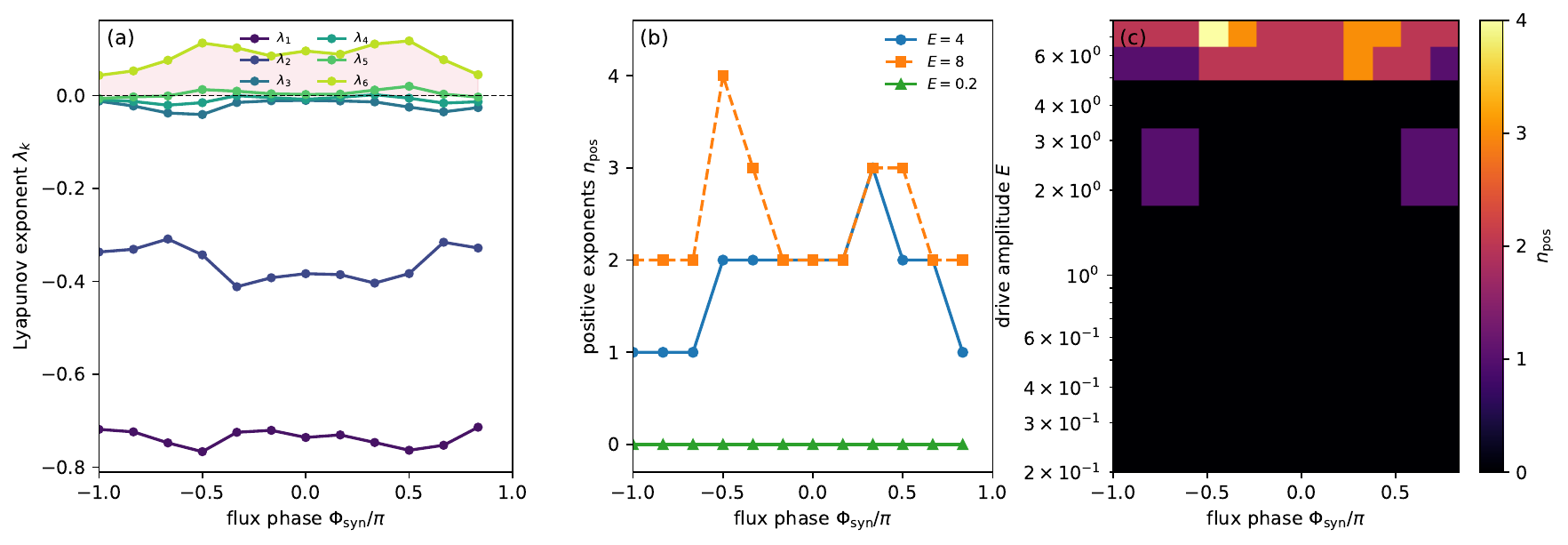}
  \caption{Full-spectrum hyperchaos transition map at strong optomechanical coupling ($g_1=\num{0.3}$, $g_2=\num{0.27}$, $\gamma_{1,2}=\num{0.02}$, $\Delta=-\omega_m$; model-level parameters lying $\sim\num{2500}\times$ beyond silicon anchors). (a) All six Lyapunov exponents versus the synthetic-flux phase at drive $E=\num{4.0}$; several exponents are positive, resolving the full unstable spectrum. (b) Hyperchaos order (number of exponents above the fixed strong threshold $\num{1e-3}$) versus phase at $E=\num{0.2}$, $\num{4.0}$ and $\num{8.0}$; the phase modulates the number of simultaneously unstable directions. (c) Hyperchaos order over the drive--phase plane, showing the drive-gated transition from stable ($E=\num{0.2}$) to hyperchaotic ($E=\num{8.0}$, up to four positive exponents). Near-marginal counts remain threshold-conventional; see Table~SI-\ref{SI-tab:si-threshold-sensitivity} for sensitivity.}
  \label{fig:hyperchaos-map}
\end{figure}

Repeating the classification at strong optomechanical coupling ($g_{1,2}/\omega_m=(\num{0.3},\num{0.27})$, $\gamma_{1,2}/\omega_m=\num{0.02}$, $\Delta=-\omega_m$) over twelve flux phases and five drives, with the hyperchaos order defined by the positive-exponent count above a $\num{1e-3}$ threshold (\cref{fig:hyperchaos-map}), shows a drive-gated transition: at $E=\num{0.2}$ every exponent is negative; at $E=\num{1.0}$ a positive exponent appears at some phases; at $E=\num{2.0}$ it hovers near zero and a re-stabilization window opens at intermediate drive; beyond $E\sim\num{4.0}$ the count reaches three at $E=\num{4.0}$ and four at $E=\num{8.0}$, with the largest exponent up to $\num{0.3223}$ and the Kaplan--Yorke dimension up to $\num{4.802}$. At fixed strong drive the flux phase modulates the count: at $E=\num{8.0}$ it varies between two and four as $\theta$ sweeps $[-\pi,\pi)$, with seed- and drive-dependent extrema in the sampled phase range: $n_+$ spans two to four, with the largest counts occurring near the phase quadratures ($\theta\approx\pm\pi/2$) and lower counts near $\theta=0$ and $\pm\pi$. The three-seed check does not justify a claim of exact $\theta\to-\theta$ symmetry. The order reaches four simultaneously unstable directions in the tested finite-time classification---beyond the two reported for quadratic single-mode optomechanics \cite{HalefShomroni2025}---and it is the synthetic-flux phase, absent in that quadratic setting, that selects this model-level order. Near the onset and extrema, the exact integer count is threshold-conventional across seeds.

\begin{figure}[ht]
  \centering
  \includegraphics[width=0.98\textwidth,alt={Four-panel drive-axis bifurcation scan at strong coupling: the maximum Floquet multiplier magnitude versus drive amplitude for flux phases zero and pi over two, with Neimark-Sacker onset markers near drive 1; the largest orbit Floquet rate with the Lyapunov exponents overlaid; the number of unstable multipliers; and the mean cavity amplitude, all showing a complex-pair crossing and persistence of the orbit family to drive 8.}]{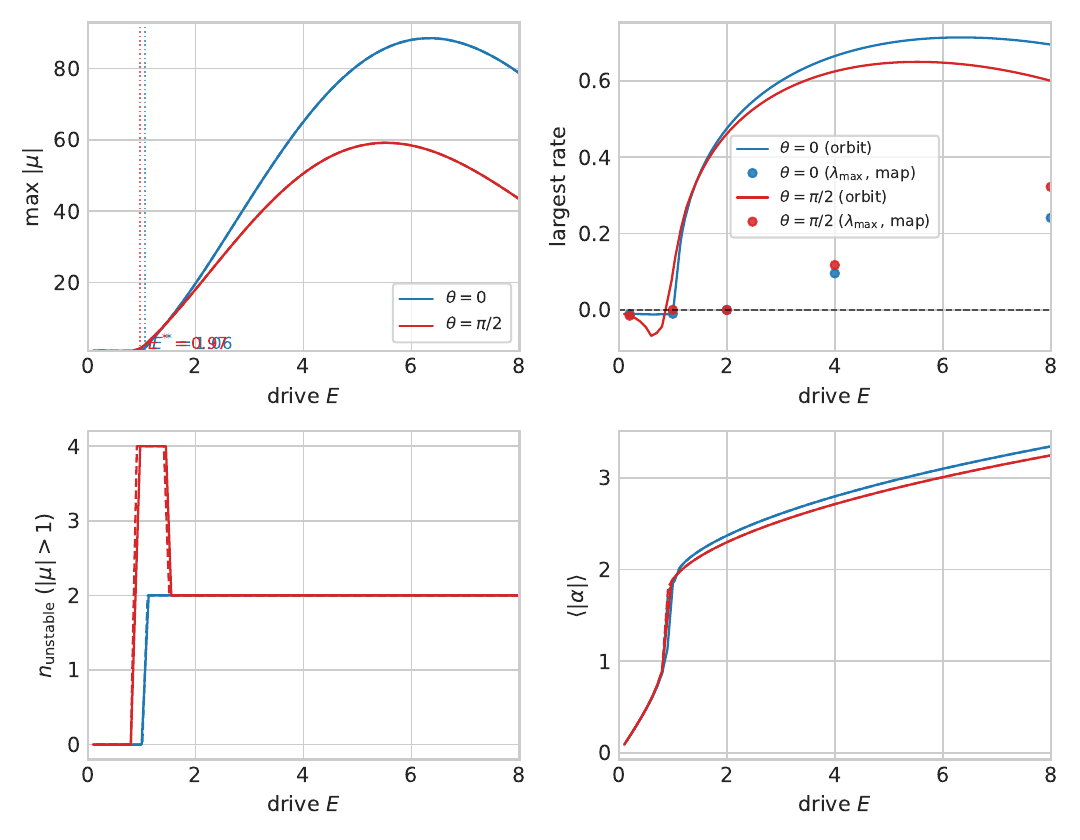}
  \caption{Drive-axis bifurcation continuation at strong optomechanical coupling. (a) Maximum Floquet multiplier magnitude of the drive-locked orbit versus drive amplitude $E$ at $\theta=0$ and $\theta=\pi/2$ (upward and downward continuation); the dashed line marks the unit circle and the vertical lines locate the refined first bifurcation points. (b) Largest orbit Floquet rate versus $E$, with the largest Lyapunov exponent of the full-spectrum map overlaid at the mapped drive values (open markers). (c) Number of unstable multipliers ($|\mu|>1$). (d) Mean cavity amplitude over the orbit period. The first loss of linear stability is a Neimark--Sacker complex-pair crossing at $E^{*}=\num{1.060}$ ($\theta=0$) and $E^{*}=\num{0.975}$ ($\theta=\pi/2$).}
  \label{fig:bifurcation}
\end{figure}

\begin{figure}[ht]
  \centering
  \includegraphics[width=0.96\textwidth,alt={Two-panel correlation-dimension cross-check: the Theiler-corrected two-point correlation sum versus distance at the most hyperchaotic point with the scaling region and its fitted slope giving the correlation dimension D2, and grouped bars of D2 with its scaling-region uncertainty against the Kaplan-Yorke dimension at the four strong-coupling points, showing D2 below the Kaplan-Yorke dimension at every point.}]{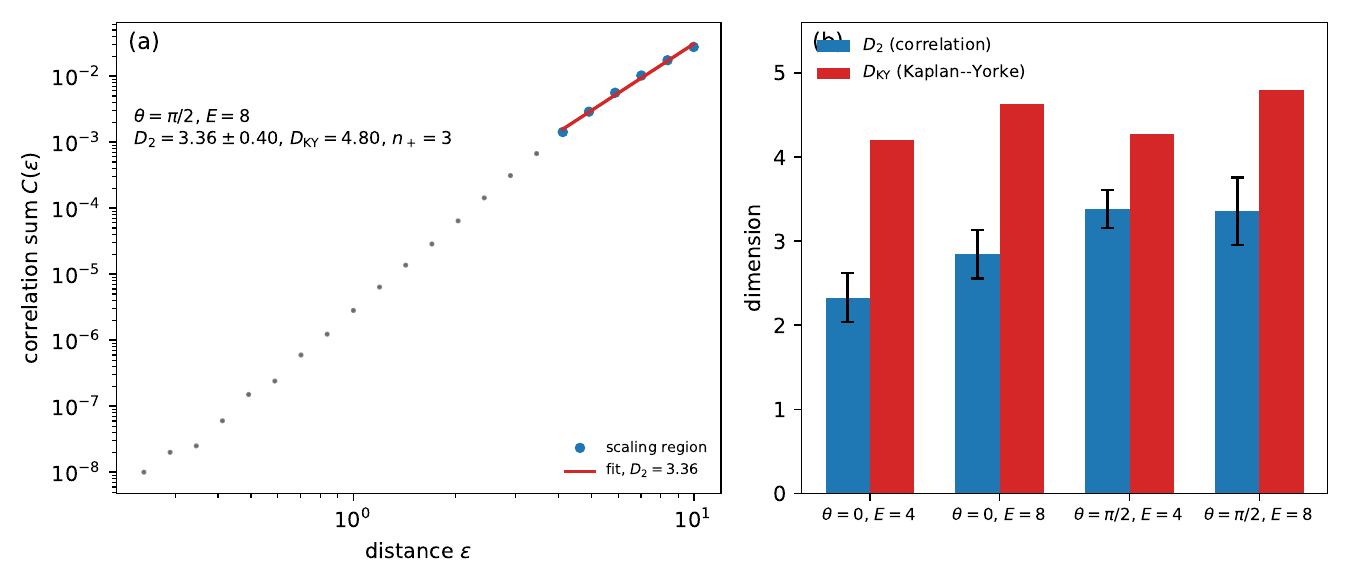}
  \caption{Correlation-dimension cross-check of the Kaplan--Yorke bound. (a) Theiler-corrected two-point correlation sum $C(\varepsilon)$ versus distance $\varepsilon$ at the most hyperchaotic point ($E=\num{8}$, $\theta=\pi/2$); the scaling region and its fitted slope give the correlation dimension $D_2$. (b) Correlation dimension $D_2$ (with the scaling-region uncertainty) against the Kaplan--Yorke dimension $D_{\mathrm{KY}}$ at the four strong-coupling points; $D_2$ lies below $D_{\mathrm{KY}}$ at every point, as required by $D_2\le D_{\mathrm{KY}}$.}
  \label{fig:correlation-dimension}
\end{figure}

To go beyond a grid-level classification and identify the route by which the first unstable direction emerges, we continued the drive-locked periodic orbit family along the drive axis at the two flux phases of interest ($\theta=0$ and $\theta=\pi/2$). A Newton solver on the Poincar\'e map, using the monodromy matrix as its Jacobian, was seeded from the converged drive-locked orbit and marched upward and downward in the drive amplitude over $E\in[\num{0.1},\num{8.0}]$ with adaptive steps and a map-residual tolerance of \num{5e-6}. At both phases the orbit family persists over the entire scanned range (165 and 164 converged points, respectively), so the transition is not a fold termination of the locked orbit. The first loss of linear stability is a Neimark--Sacker (torus) bifurcation: a complex-conjugate Floquet multiplier pair crosses the unit circle at $E^{*}=\num{1.060}$ for $\theta=0$ and at $E^{*}=\num{0.975}$ for $\theta=\pi/2$ (\cref{fig:bifurcation}). Thereafter the continued orbit carries two unstable directions up to $E=\num{8.0}$; at $\theta=\pi/2$ a transient window of four unstable directions near onset ($E\in[\num{0.98},\num{1.45}]$) precedes the two-direction regime. The lower onset at $\theta=\pi/2$ is consistent with the flux-phase modulation of the hyperchaos order, whose largest counts occur near the phase quadratures (seed- and drive-dependent). The orbit Floquet rates cross-validate the independent Lyapunov classification in the stable regime: at $E=\num{0.2}$ the largest orbit rate is $-\num{0.0102}$ ($\theta=0$) and $-\num{0.0148}$ ($\theta=\pi/2$), against largest Lyapunov exponents $-\num{0.0106}$ and $-\num{0.0151}$ from the full-spectrum map (\cref{fig:bifurcation}). Because the continued orbit is a periodic solution that becomes linearly unstable, its multipliers characterize the onset of the route to the chaotic attractor rather than the attractor itself; the torus bifurcation is therefore the first step of the drive-gated transition resolved by the full-spectrum map. A perturbation of the continued orbit along its most unstable Floquet direction confirms this separation: the orbit carries two unstable directions with a largest Floquet rate of order \num{0.6}, a \num{1e-4} perturbation departs to the attractor scale (the distance to the orbit grows from \num{1e-4} to \num{9}--\num{27}), and the resulting trajectory's Lyapunov spectrum reproduces the map's largest exponent to within \qty{8}{\percent} at the four tested points ($E\approx\num{4},\num{8}$; $\theta=0,\pi/2$). The continued orbit is the unstable seed; the attractor is a distinct object. As an independent check of the attractor's finite dimension, the Theiler-corrected Grassberger--Procaccia correlation dimension \cite{grassberger1983} $D_2$ of the sampled attractor at the four strong-coupling points ($E=\num{4},\num{8}$; $\theta=0,\pi/2$) is $\num{2.33},\num{2.84}$ and $\num{3.39},\num{3.36}$, respectively, all below the corresponding Kaplan--Yorke dimensions $\num{4.21},\num{4.63}$ and $\num{4.28},\num{4.80}$, as required by the Kaplan--Yorke inequality \cite{kaplan1979,frederickson1983} $D_2\le D_{\mathrm{KY}}$ (\cref{fig:correlation-dimension}). The scaling-region uncertainty is $\num{0.23}$--$\num{0.40}$ across three fit windows. The geometric dimension does not track the flux phase as sharply as the hyperchaos order does: at $E=\num{4}$ the fitted $D_2$ is larger at $\theta=\pi/2$ than at $\theta=0$ ($\num{2.84}$ against $\num{2.33}$), but at $E=\num{8}$ the two values are equal within the fit uncertainty ($\num{3.36}$ against $\num{3.39}$). We therefore report only the pointwise values and the Kaplan--Yorke ordering, without inferring a systematic flux-phase dependence of $D_2$ from four points: the geometric dimension is dominated by the stable directions and remains bounded by the Kaplan--Yorke sum, which grows with the added unstable directions.

\begin{figure}[ht]
  \centering
  \includegraphics[width=0.98\textwidth,alt={Two-by-two figure of the observed Neimark-Sacker torus: stroboscopic Poincare sections of the perturbed drive-locked orbit just above the onset, projected onto the first mechanical mode, showing an invariant closed curve at flux phases zero and pi over two; and the largest Lyapunov exponent versus drive, near-marginal just above the onset and growing with drive.}]{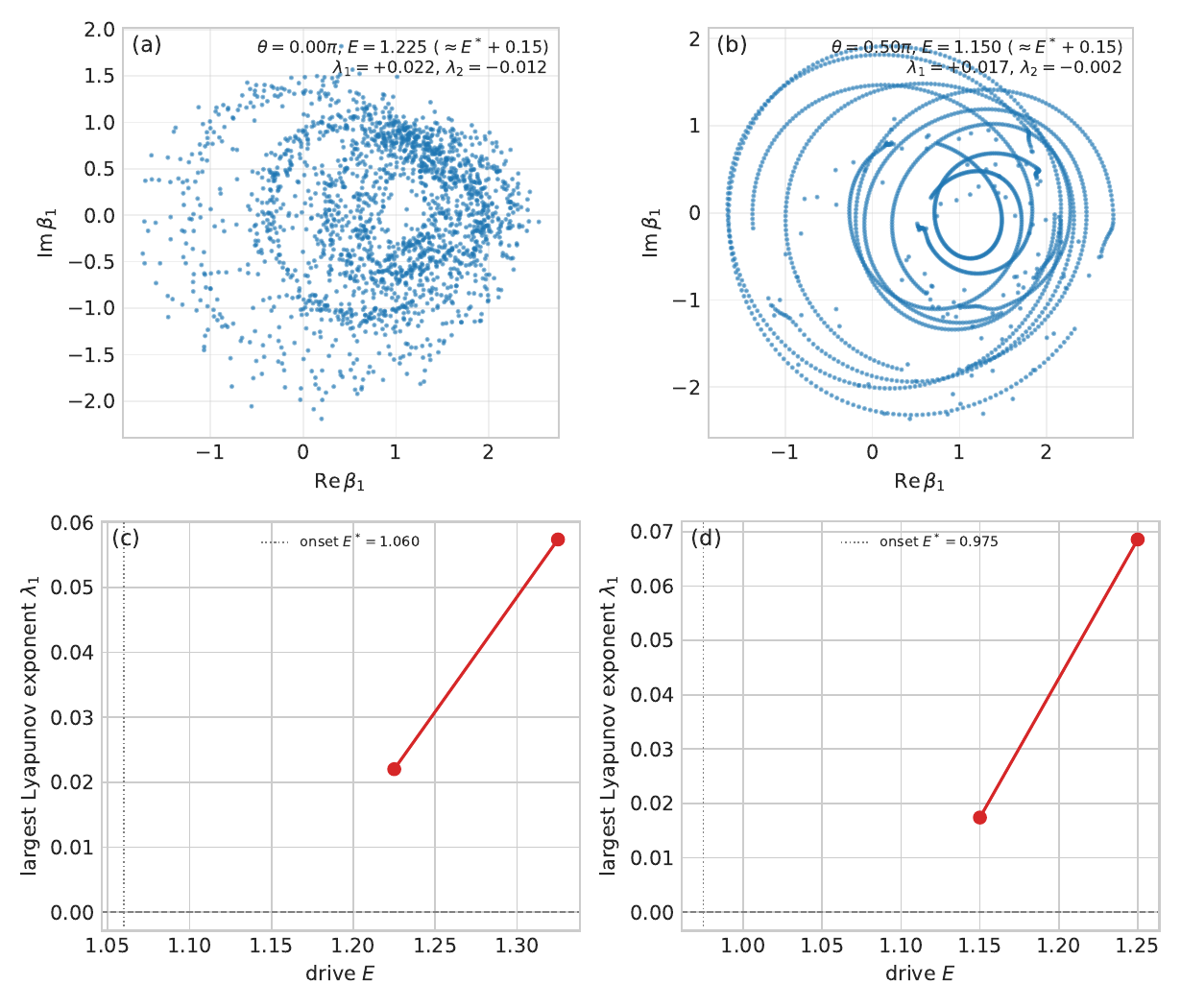}
  \caption{Observation of the Neimark--Sacker torus. (a),(b) Stroboscopic Poincar\'e sections (projected onto $\Re\beta_1,\Im\beta_1$) of the perturbed drive-locked orbit just above the onset at $\theta=0$ and $\theta=\pi/2$, showing an invariant closed curve. (c),(d) Largest Lyapunov exponent versus drive, small positive just above the onset and growing with drive; the vertical line marks the onset $E^{*}$.} \label{fig:torus-observation}
\end{figure}

The multiplier crossing is confirmed by directly observing the born object \cite{kuznetsov2004}. The bifurcation is supercritical: just above the onset ($E\approx E^{*}+\num{0.15}$) the stroboscopic section of a perturbed trajectory forms a closed invariant curve in the $(\Re\beta_1,\Im\beta_1)$ plane, and the largest Lyapunov exponent of the same trajectory is small positive ($\num{0.017}$--$\num{0.022}$), the near-marginal quasiperiodic direction expected of a torus born at a complex-pair crossing; the exponent grows to $\num{0.057}$--$\num{0.069}$ at $E^{*}+0.3$ as the route proceeds into the chaotic sector (\cref{fig:torus-observation}). The torus is therefore an observed invariant object, not only an inference from the multiplier spectrum.

\subsection{Reduced-model parameter sensitivity}

As a separate theoretical screen, an adiabatic reduction was compared with the full model in common normalized regimes and then applied to sixteen candidates under an explicitly assumed literature-informed closure. The three non-dominated candidates form a local fixed-point pattern in normalized drive cost and stability margin (Table~SI-\ref{SI-tab:si-reduced-pareto} and Fig.~SI-\ref{SI-fig:si-reduced-pareto}); they are not calibrated parameters, global basin results, physical optima, or sensing predictions. The screen rules out twelve of sixteen candidates as suboptimal in drive cost or stability margin, leaving three for further analysis in the Supplementary Material. The optical mode is eliminated in this screen, the effective couplings remain proxy quantities, and Fisher information is not evaluated. Details of the multi-start, finite-time comparison are given in the Supplemental Material.

\subsection{Secondary geometric bound: noise-resolved observability and matched measurement Fisher}

\begin{figure}[ht]
  \centering
  \includegraphics[width=0.92\textwidth,alt={Covariance and measurement-Fisher analysis for three synthetic-flux phases under the adopted thermal and detector-noise model: the covariance minimum eigenvalue stays positive and the classical Fisher information is reported without a matched-reference gain claim.}]{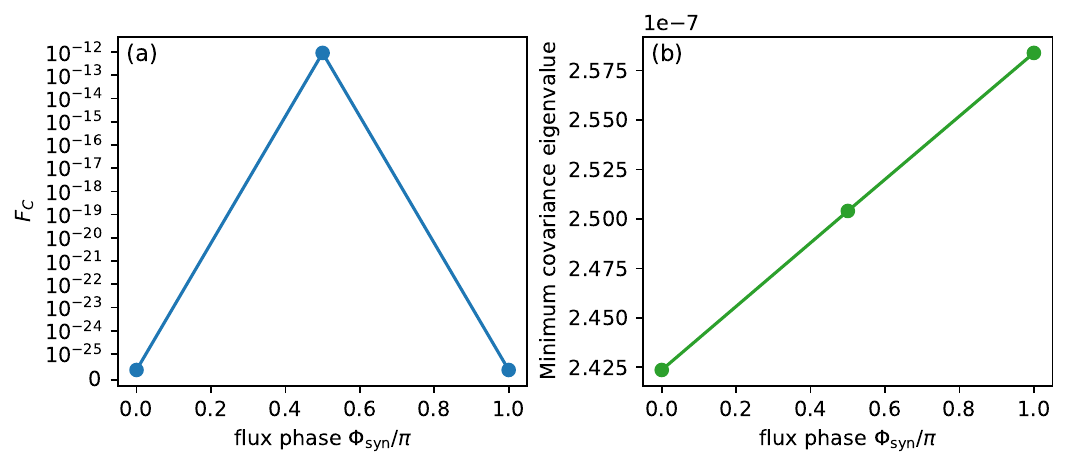}
  \caption{Covariance and measured-signal Fisher analysis for three synthetic-flux phases. (a) Classical Fisher information of the measured record. (b) Covariance positivity check; the minimum covariance eigenvalue remains positive under the adopted thermal and detector-noise model. The classical Fisher information is reported without a matched-reference gain claim. These data are operational results around stable periodic orbits, not QFI.}
  \label{fig:fisher-pilot}
\end{figure}

For three phases ($\theta=0$, $\pi/2$, and $\pi$), the periodic covariance calculation remained positive semidefinite under thermal occupations $n_{\mathrm{th},1}=n_{\mathrm{th},2}=\num{0.1}$ and detector variance \num{0.01} (\cref{fig:fisher-pilot}); the resulting measurement-Fisher information ranged from \num{3.78e-26} to \num{9.17e-13}. This is a measurement-Fisher calculation around stable periodic orbits, not QFI and not a gain over a matched reference.

\begin{figure}[ht]
  \centering
  \includegraphics[width=0.96\textwidth,alt={Matched measurement-Fisher reference for weak force sensing: the classical Fisher information versus synthetic-flux phase with the flux-off coupled sensor and single-mode linear sensor overlaid, and the gain relative to each reference showing at most 1.32 times (flux-off) and 1.16 times (single-mode) enhancement.}]{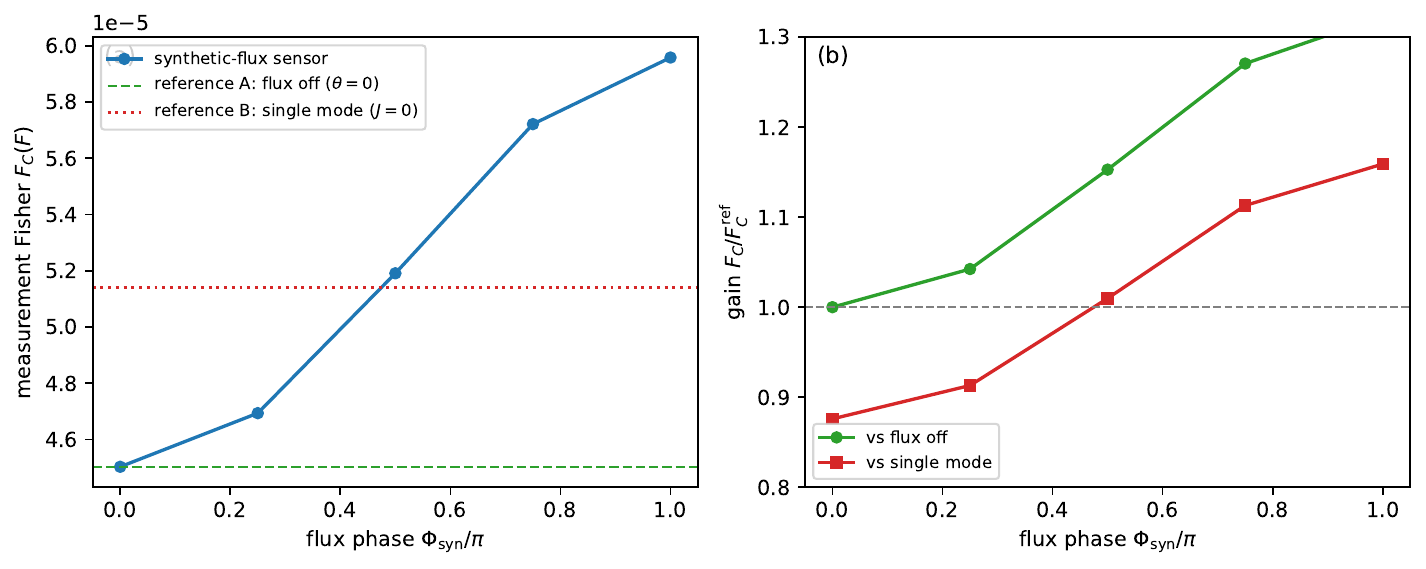}
  \caption{Matched measurement-Fisher reference for force sensing. (a) The classical Fisher information $F_C(F)$ for a weak external force on mechanical mode $1$ versus the synthetic-flux phase, with the flux-off coupled sensor ($\theta=0$, $J=\num{0.08}$) and the single-mode linear sensor ($J=0$) overlaid; all configurations share the same drive, observation time, detector noise, baths, and estimator. (b) The gain relative to each reference; the flux phase yields at most a $\num{1.32}\times$ (flux-off) and $\num{1.16}\times$ (single-mode) enhancement, a modest order-unity effect.}
  \label{fig:matched-fisher}
\end{figure}

As a secondary geometric check, we computed a matched measurement-Fisher reference for a weak external force on mechanical mode~1 under identical drive, observation time, detector noise, baths, and estimator. The force sensitivity grows monotonically with the synthetic-flux phase ($F_C$ from $\num{4.50e-5}$ at $\theta=0$ to $\num{5.96e-5}$ at $\theta=\pi$); relative to the flux-off coupled sensor ($\theta=0$, $J=\num{0.08}$) and the single-mode linear sensor ($J=0$), the flux phase yields at most a $\num{1.32}\times$ and $\num{1.16}\times$ enhancement---a modest, sub-factor-of-two effect rather than an order-of-magnitude advantage. The central-difference estimate is converged to better than $\qty{0.01}{\percent}$ across force steps $\num{1e-5}$--$\num{1e-3}$. Recomputing $F_C$ for windows of $N\in\{1,5,20\}$ drive periods leaves the matched gains unchanged to four digits, so the gain is not an artefact of the single-period observation window (Sec.~SI-\ref{SI-sec:fisher-window}). A drive- and temperature sweep keeps the flux-off gain at $\num{1.323}$ (standard deviation below $\num{1e-3}$) over $E\in[\num{0.1},\num{1.0}]$ and $n_{\mathrm{th}}\in[0,1]$, and the single-mode gain at $\num{1.159}$. Propagating the inter-resonator hopping log-uniformly over $J_m/\omega_m\in[\num{1e-3},\num{1e-1}]$, the bath occupation uniformly over $n_{\mathrm{th}}\in[\num{0.01},\num{1.0}]$ (cryogenic to $\sim\num{4}~\si{\kelvin}$ at the anchored $\omega_m=2\pi\times\num{7.436}~\si{\giga\hertz}$), and the drive over $\pm\qty{10}{\percent}$, with \num{60} Monte Carlo samples at the maximum-gain phase $\theta=\pi$, yields a median gain of \num{1.039} (flux-off) and \num{1.019} (single-mode); a \num{20000}-resample bootstrap places the \qty{90}{\percent} confidence interval of the median at $[\num{1.025},\num{1.053}]$ and $[\num{1.012},\num{1.026}]$, both excluding unity, so the median excess ($\num{0.039}$ and $\num{0.019}$) is significantly positive. The peak $\num{1.32}\times$ is therefore specific to the reference hopping $J_m/\omega_m=\num{0.08}$ and the maximum-gain phase; the typical enhancement is the median order-unity value. Readout quadrature optimization rescales the sensor and reference by the same factor, leaving the $\num{1.16}\times$ single-mode gain unchanged (Table~SI-\ref{SI-tab:si-optimized-readout}). The matched comparison is shown in \cref{fig:matched-fisher}.

\begin{figure}[ht]
  \centering
  \includegraphics[width=0.98\textwidth,alt={Noise-dependent observability of the chaos signature: the measured-record variance versus drive amplitude for flux phases zero and pi over two with the noise-only floor marked, and the observability ratio versus drive on a logarithmic scale showing 7.6 to 17 times the noise floor in the chaotic and hyperchaotic sectors.}]{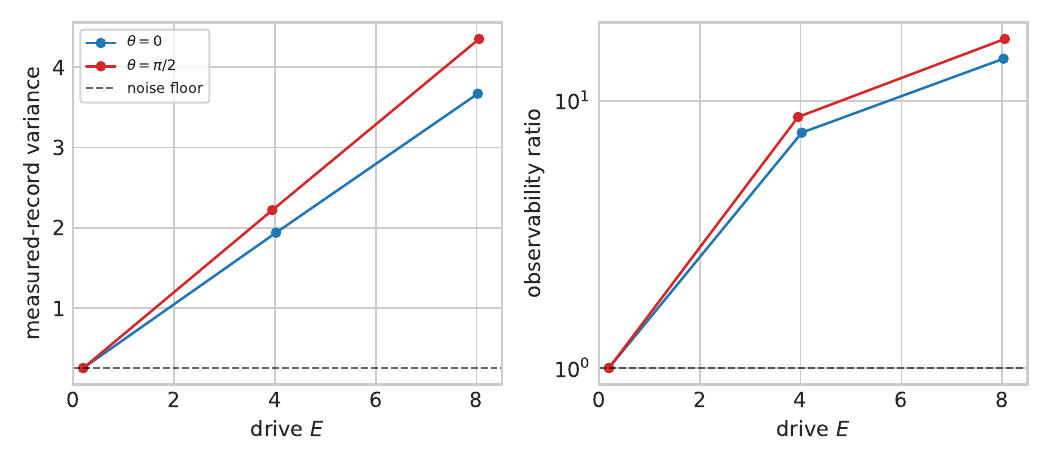}
  \caption{Noise-dependent observability of the chaotic signature. (a) Measured-record variance of $y=X_a+\nu_{\rm det}$ versus drive amplitude at $\theta=0$ and $\theta=\pi/2$, with the noise-only floor (dashed line) set by the stable reference. (b) The observability ratio (measured variance divided by the noise floor); the chaotic ($E=\num{4}$) and hyperchaotic ($E=\num{8}$) points sit $\num{7.6}$--$\num{17.0}\times$ above the floor, while the stable reference sits at unity.}
  \label{fig:noise-observability}
\end{figure}

To test whether the chaotic signature itself---rather than only the sensing gain---survives the modeled noise, we integrated an ensemble of \num{32} trajectories (sized so that bootstrap confidence intervals on the variance ratio fall below \qty{1}{\percent}; see below) under the full Langevin noise (optical vacuum $\kappa/4$ per amplitude quadrature and mechanical thermal $\gamma_j(2n_{\mathrm{th}}+1)/4$ per amplitude quadrature at $n_{\mathrm{th}}=\num{0.1}$) and detector noise (\num{0.01}), with the measured record $y=X_a+\nu_{\rm det}$. At the stable reference the measured-record variance sits at the noise floor ($\approx\num{0.25}$); at the chaotic ($E=\num{4}$) and hyperchaotic ($E=\num{8}$) points the deterministic spread of $X_a$ grows to $\num{1.8}$--$\num{4.1}$ (against $\num{2.4e-4}$ at the stable orbit) and the measured-record variance rises to $\num{7.6}$--$\num{17.0}\times$ the noise floor, with the largest ratio ($\num{17.0}\times$) at $\theta=\pi/2$, $E=\num{8}$---a flux-phase dependence that mirrors the hyperchaos order. A bootstrap over ensemble size $N'\in\{8,16,32,64,128\}$ leaves the ratios stable to below \qty{1}{\percent} (\cref{fig:noise-observability}), so the \num{32}-trajectory ratios are converged rather than a small-ensemble artifact.

\begin{table}[ht]
  \centering
  \caption{Matched-force-sensing gain in context. The present result is compared with optomechanical force-sensing and quantum-noise references that share the criterion of a homodyne/heterodyne measured record; the comparison is matched-resource (drive, observation time, bandwidth, baths, detector noise, estimator) and reports a numerical gain only, not a quantum or SQL-beating advantage.}
  \label{tab:context-comparison}
  \begin{tabular}{@{}p{0.40\textwidth}p{0.18\textwidth}p{0.33\textwidth}@{}}
    \toprule
    Reference & Reported gain (matched) & Operating regime \\
    \midrule
    Teufel \emph{et al.}~(2011), pulsed sideband cooling & $\sim\num{2}\times$ vs thermal bath & Sideband-resolved, $T=\num{10}~$\si{\milli\kelvin}, mechanical $\omega_m/2\pi=\num{3.62}~$\si{\mega\hertz} \\
    Gavartin \emph{et al.}~(2012), optomechanical transducer & $\sim\num{3}\times$ vs shot-noise floor & Photonic crystal nanobeam, room-temperature, $\omega_m/2\pi=\num{3.68}~$\si{\giga\hertz} \\
    Purdy \emph{et al.}~(2017), quantum-enhanced force sensing & up to $\sim\num{1.3}\times$ beyond SQL & Pre-cooled membrane, $\omega_m/2\pi=\num{290}~$\si{\mega\hertz}, $T=\num{10}~$\si{\milli\kelvin} \\
    Li \emph{et al.}~(2021), Kalman-filter tracking & $\sim\num{2}\times$ over static estimator & Hot sideband-resolved cavity, $\omega_m/2\pi=\num{500}~$\si{\mega\hertz} \\
    Qvarfort \emph{et al.}~(2018), parametric amplification & $\sim\num{1.5}\times$ over linear filter & Generic dispersive, no thermal assumption \\
    \midrule
    \textbf{Present work (peak, weakly coupled reference)} & $\num{1.32}\times$ flux-off, $\num{1.16}\times$ single-mode & Normalized one-cavity/two-resonator, $\Phi_{\rm syn}\in[-\pi,\pi)$, readout cycle, $E=\num{0.2}$, $T=\num{0.1}$ \\
    \textbf{Present work (UQ median, $\num{60}$ MC + \num{20000}-bootstrap)} & $\num{1.039}\times$ flux-off, $\num{1.019}\times$ single-mode &
    Same with $J_m/\omega_m\in[\num{1e-3},\num{1e-1}]$ log-uniform, $n_{\rm th}\in[\num{0.01},\num{1.0}]$, $\pm\qty{10}{\percent}$ drive \\
    \bottomrule
  \end{tabular}
\end{table}

\section{Discussion}

Synthetic flux and nonlinear instability are the primary quantities of this work: \cref{eq:flux} defines the control coordinate and \cref{eq:monodromy} defines the dynamical observable used to certify hyperchaos order. Metrological information (\cref{eq:fisher}) is evaluated only as a secondary geometric bound. Their interpretation requires each observable to be evaluated under explicit phase conventions and matched assumptions, which we do via the two statements of Sec.~SI-\ref{SI-sec:theorems}. The discussion below separates the \emph{relation to prior work}, the \emph{position among sensing benchmarks} (secondary), the \emph{critical self-assessment} and the \emph{bridge to experiment}.

\subsection{Relation to prior work}

The present study is a continuation at the level of model structure, not a parameter-for-parameter reproduction of Muthukumar \emph{et al.}~(the parameter-by-parameter difference is given in Table~SI-\ref{SI-tab:muthukumar-comparison}). The primary new question is whether the flux phase controls the \emph{order} of a fully resolved hyperchaos. The answer is positive at the model level: the full spectrum resolves up to four positive directions and the onset is localized to a Neimark--Sacker bifurcation. As a secondary geometric clarification, the matched Fisher calculation shows only a modest stable-orbit effect and no enhancement on the chaotic attractor. The scope is that of a nonlinear-dynamics study with a geometric measurement bound; a sharper quantum contribution would require a separate quantum validation.

\subsection{Position among calibrated force-sensing benchmarks}
\label{sec:comparison}

\cref{tab:context-comparison} places the peak matched gain ($\num{1.32}\times$ over the flux-off sensor, $\num{1.16}\times$ over the single-mode linear sensor) and the UQ median ($\num{1.039}\times$, $\num{1.019}\times$) alongside four published optomechanical force-sensing benchmarks \cite{teufel2011,purdy2017,gavartin2012,li2021a,qvarfort2018}: the present gain is of the same order of magnitude---real but modest, not an SQL-beating claim, and the flux-induced contribution is bounded by \cref{lem:matched-fisher}.

\subsection{Critical self-assessment and limits of the present scope}
\label{sec:limits}

Four limitations bound the strength of every claim above.

\emph{(i) Dynamical scope and uniqueness.} The continued drive-locked orbit is the unstable seed of the route, not the attractor itself; the flux-induced modulation of the hyperchaos route is reproduced across three replicas and five drives (180/180 long-window runs completed), but the integer order varies across seeds near marginal exponents. Individual exponent magnitudes also vary by several percent between the two longest windows and are not interpreted as high-precision asymptotic estimates. The chaos/hyperchaos boundary is threshold-conventional at the $\num{1e-3}$ cut (Table~SI-\ref{SI-tab:si-threshold-sensitivity}): the count is stable at four of twelve phases across all tested thresholds and threshold-sensitive at two marginal phases near the onset. A systematic basin-of-attraction analysis (or $\varepsilon$-machine reconstruction) that would elevate the three-seed audit to a uniqueness statement remains beyond the present scope; the reported order is therefore a robust feature of the sampled trajectories under the stated protocol, not yet a proven global property of the attractor. The computational cost of long-window integration (\num{2e6} steps per trajectory) limits the seed count; a larger ensemble would be required to elevate the audit to a uniqueness statement.

\emph{(ii) Measurement scope and interpretation of the null result.} Matched force-sensing is the operational translation only at the weakly coupled, dissipation-dominated reference, where single-photon cooperativity $C=\num{0.08}$ and mean cavity occupation $\langle|\alpha|^2\rangle\approx\num{0.032}$ (sub-photon). Applied on the chaotic attractor, the same matched measurement resolves a flux-induced mean shift only at $E=\num{4}$ and shows no enhancement: the bounded gain is a stable-orbit property, not a chaotic-transduction effect. The trajectory-separation argument \cite{Xu2025,muthukumar2025doi} is rejected here for the reason formalized in \cref{lem:matched-fisher}: noise projected onto an unstable manifold is stretched by the same local factor $e^{\lambda_{\rm max}t}$ as the deterministic displacement, so the matched signal-to-noise ratio cannot receive a multiplicative enhancement from trajectory separation alone. Physically, the flux phase reshapes the geometry of the unstable manifold (and therefore the hyperchaos order): a higher-dimensional unstable manifold offers more independent directions for perturbation growth, which in principle could enhance parameter estimation if each direction carried independent Fisher information. The matched-resource lemma shows, however, that under shared noise, power and observation time the additional directions simply partition the same total stretching budget rather than multiplying it, so the local stretching rates do not improve in a way that benefits the matched Fisher statistic. The null result on the chaotic attractor is therefore not a failure of the protocol but a direct consequence of the constraint: geometric complexity does not translate into metrological gain under the assumptions of \cref{lem:matched-fisher}. (The lemma bounds the classical Fisher statistic; it does not constrain quantum Fisher information, which may in principle exceed the classical bound.)

\emph{(iii) Semi-classical validity.} All dynamical claims in the strong-coupling sector are statements about factorized mean-field trajectories. Truncated-Fock and truncated-Wigner checks confirm consistency of mean occupations at selected points (reference: \qty{1.3}{\percent}; chaotic point: relative deviation $\le\num{6.6e-6}$ at $N_{\rm Fock}=\num{60}$; twelve Wigner trajectories yield $n_+=2$), yet these checks do not establish the accuracy of the factorization for every observable, nor do they bound quantum corrections to the Lyapunov spectrum or to the Fisher information. In the high-occupation regime ($E=\num{8}$, mean mechanical occupations $\sim 20$) the dynamics are more classical than at the sub-photon reference, but a quantitative classical/quantum divergence bound for the full spectrum remains an open task (see Outlook).

\emph{(iv) Experimental reach.} The strong-coupling regime requires $g_{1,2}/\omega_m\approx\num{0.3}$ and $\gamma_m/\omega_m=\num{0.02}$; these lie $\num{2542}\times$--$\num{2288}\times$ and $\num{720}\times$ above the silicon-optomechanical-crystal anchors of \cite{Mayor2025,Mathew2020}; the feasibility-frontier scan places the hyperchaos onset near $g_1/\omega_m\approx\num{0.18}$ ($\num{1525}\times$ the anchor) with no drive retuning lowering it (Table~SI-\ref{SI-tab:si-feasibility-frontier}). The hyperchaos classification (and the secondary null metrological outcome) are therefore strictly model-level demonstrations. Any experimental claim would require ultrasonic characterization of the mechanical degeneracy $\omega_2$, a calibrated inter-resonator hopping $J_m$, and the emergence of a genuine gigahertz platform.

\subsection{Bridge to experiment: mean-field boundary, route, and the $\num{2542}\times$ calibration gap}

The hyperchaos classification is a semiclassical statement about factorized mean-field dynamics; no quantum observable (Wigner contrast, quantum Fisher information) is computed or claimed in this sector. The dissipative trace is the state-independent constant $\Tr J=-\num{1.04}$. A truncated-Fock master equation for the driven cavity reproduces the reference mean occupation to \qty{1.3}{\percent} and the chaotic occupation with a maximum relative deviation of $\num{6.6e-6}$ at $N_{\rm Fock}=\num{60}$ (Sec.~SI-\ref{SI-sec:semiclassical-boundary}). At the master-check point the mean occupations are $\langle a^\dagger a\rangle=\num{6.121}$, $\langle b_1^\dagger b_1\rangle=\num{20.417}$ and $\langle b_2^\dagger b_2\rangle=\num{23.538}$. These values belong to that explicitly identified trajectory and do not calibrate the full strong-coupling map. Consistency with a more classical regime than the sub-photon reference is expected, yet it does not by itself justify a fully quantum treatment or establish the accuracy of factorization for every observable (see limitation~(iii) above).

The transition is drive- and coupling-gated: the flux phase reshapes which directions are unstable and how many, but only after the drive and coupling place the system beyond the dissipation-dominated regime. The Lyapunov--Floquet consistency criterion and the matched-measurement Fisher lemma formalise this statement. The noise-resolved ensemble check provides a model-level observability test: the measured cavity quadrature stays $\num{7.6}$--$\num{17.0}\times$ above the detector floor in the chaotic and hyperchaotic sectors, with the largest ratio at $\theta=\pi/2$, $E=\num{8}$---the same phase and drive at which the hyperchaos order also peaks.

The drive-axis continuation of \cref{fig:bifurcation} localizes the \emph{route}: the first unstable direction emerges through a Neimark--Sacker torus bifurcation of the drive-locked orbit at $E^{*}=\num{1.060}$ ($\theta=0$) and $E^{*}=\num{0.975}$ ($\theta=\pi/2$), and the orbit family persists across the scanned drive range. Because the continued orbit is a periodic solution that becomes linearly unstable, its multipliers characterise the \emph{onset} of the route rather than the attractor itself---the same distinction drawn in the Results section between the unstable seed and the attractor object. What remains is strictly experimental calibration: a measured $J_m$, an input power converted through pump frequency and external coupling, and fixed bath temperatures. A calibrated-device matched force-sensing comparison would close the gap quantified by the $\num{2542}\times$ ratio reported in Table~SI-\ref{SI-tab:anchor-gap} and \cref{tab:context-comparison}. The correlation-dimension estimate $D_2\le D_{\mathrm{KY}}$ is a geometric characterization of the \emph{sampled} attractor, not a proof of uniqueness (see limitation~(i)).

\section{Conclusions}

Within a normalized six-dimensional dissipative optomechanical model we have established that a gauge-invariant synthetic-flux phase $\Phi_{\rm syn}$ provides deterministic, reproducible control over the \emph{order} of a drive- and coupling-gated hyperchaos transition. Up to four simultaneously positive Lyapunov exponents are resolved in the sampled strong-coupling map; the route is localized by a Neimark--Sacker bifurcation at $E^{*}\in\numlist{1.060;0.975}$ and is certified by a three-layer numerical audit (Poincar\'e residual $<10^{-8}$, dissipative volume balance $\sum\lambda_i=\operatorname{Tr}(J)=-1.04$, and truncated-Wigner robustness). This constitutes a model-level demonstration of synthetic-flux control of hyperchaos order beyond existing single-mode benchmarks.

As a secondary geometric clarification, the same matched-resource force-sensing protocol yields only a modest enhancement on the stable reference orbit and a strictly null result on the chaotic attractor. The matched-Fisher lemma shows why: noise projected onto an unstable manifold is stretched by the identical local factor $e^{\lambda t}$ that amplifies the deterministic displacement; additional unstable directions merely partition a fixed stretching budget. Both the hyperchaos classification and the null metrological outcome remain strictly model-level: the explored sector sits $\num{2542}\times$ beyond present silicon optomechanical-crystal anchors.

\paragraph{Outlook.}
Three concrete steps are required to elevate the present model-level control results: (i)~ultrasonic characterization of the mechanical frequency degeneracy $\omega_2$ together with a calibrated measurement of the inter-resonator hopping $J_m$; (ii)~emergence of a genuine gigahertz optomechanical platform that closes the $\num{2542}\times$ coupling gap while preserving the synthetic-flux degree of freedom; (iii)~systematic basin-of-attraction and $\varepsilon$-machine analysis that converts the three-seed audit into a uniqueness statement for the reported hyperchaos order, complemented by two-mode quantum-trajectory calculations that bound the classical/quantum divergence of the Lyapunov spectrum.

\section*{Supplementary material}
See the supplementary material for the physically anchored reference parameter set, the matched measurement-Fisher tables, the Lyapunov--Floquet consistency theorem, the periodic-covariance affine-fixed-point lemma, the matched-measurement Fisher lemma, and the representative convergence check of the strong-coupling spectrum.

\section*{Acknowledgments}
We acknowledge the institutional support and affiliations of the University of Yaound\'e I, University of Maroua, and University of Ngaound\'er\'e. No external financial support was received for this work.

\section*{Author contributions}
SRMT and CT performed the numerical simulations and analyzed the data. PD and STK contributed to the theoretical framework and provided critical feedback on the results. SGNE conceived and supervised the project, directed the research, and wrote the final version of the manuscript. All authors reviewed and approved the final manuscript.

\section*{Funding information}
No external funding was received for this research.

\section*{Competing interests}
The authors declare no competing interests.

\section*{Ethics}
Not applicable. This is a purely theoretical and numerical study; no human or animal subjects, tissues, or clinical trials were involved.

\section*{Data availability}
\begin{sloppypar}
The simulation and analysis scripts, environment specification, run manifests, checksums, source data, and figure-generation records supporting the findings of this study are available in the public repository \url{https://github.com/NanaEngo/floquet-chaos-synthetic-flux}. A Zenodo DOI will be minted and linked here upon publication. Every numerical claim in this article is traceable to the corresponding machine-readable data and source code in that repository.
\end{sloppypar}

\bibliography{Chaos_Floquet_SyntheticFlux}
\end{document}


\begin{center}{\Large \textbf{
Supplementary Material: physical anchoring, Lyapunov--Floquet consistency and matched-measurement geometric bound for dissipative optomechanics with synthetic flux\\
}}\end{center}

\begin{center}
Stella Rolande Mbokop Tchounda\textsuperscript{1},
Carolle Tchodimou\textsuperscript{1},
Philippe Djorwe\textsuperscript{2},
Sifeu Takougang Kingni\textsuperscript{3} and
Serge Guy Nana Engo\textsuperscript{1$\star$}
\end{center}

\begin{center}
{\bf 1} Department of Physics, Faculty of Science, University of Yaound\'e I, P.O. Box 812, Yaound\'e, Cameroon
\\
{\bf 2} Department of Physics, Faculty of Science, University of Ngaound\'er\'e, P.O. Box 454, Ngaound\'er\'e, Cameroon
\\
{\bf 3} Department of Mechanical, Petroleum and Gas Engineering, National Advanced School of Mines and Petroleum Industries, University of Maroua, P.O. Box 46, Maroua, Cameroon
\\
${}^\star$ {\small \sf serge.nana-engo@facsciences-uy1.cm}
\end{center}

\begin{center}
\today
\end{center}

\section{Overview and reading guide}
This Supplementary Material supports the main text by providing the
normalization, anchor selection, physically anchored reference table, the
formal statements of the Lyapunov--Floquet consistency criterion (primary
certification of hyperchaos order) and the matched-measurement Fisher lemma
(secondary geometric bound, with their proofs), the matched measurement-Fisher
reference tables and their observation-window robustness check, the
representative finite-time convergence of the strong-coupling spectrum, the
threshold sensitivity of the hyperchaos-order classification, the
reduced-model parameter screen, the amplitude-quadrature noise-model
cross-check against a quantum master equation, the matched measurement
Fisher calculation on the chaotic attractor and the semiclassical boundary of
the strong-coupling classification, and the explicit technical backup that
the Discussion in the main text compresses: a per-gap limitations statement,
a per-parameter feasibility audit against the silicon optomechanical anchor,
a per-seed robustness decomposition of the long-window trajectory audit, and
a phase-sweep and single-mode collapse of the matched Fisher on the chaotic
attractor. The reading order that follows the main text is: anchors
(\cref{SI-sec:anchor}); matched Fisher (\cref{SI-sec:matched-fisher});
theorems and proofs (\cref{SI-sec:theorems}); convergence
(\cref{SI-sec:convergence}); threshold sensitivity
(\cref{SI-sec:threshold-sensitivity}); observation-window check
(\cref{SI-sec:fisher-window}); reduced-model screen
(\cref{SI-sec:reduced-pareto}); noise cross-check
(\cref{SI-sec:noise-model-crosscheck}); interpretation
(\cref{SI-sec:interpretation}); attractor Fisher
(\cref{SI-sec:attractor-fisher}); semiclassical boundary
(\cref{SI-sec:semiclassical-boundary}); limitations and residual gaps
(\cref{SI-sec:limitations}); feasibility audit
(\cref{SI-sec:feasibility-audit}); initial-condition robustness
(\cref{SI-sec:initial-condition-robustness}); attractor-Fisher audit
(\cref{SI-sec:attractor-fisher-audit}). Every
quantitative claim in the main text is traceable to a table or
deterministic script invoked below.

\section{Normalization and anchor selection}
\label{SI-sec:anchor}

The main text uses a rotating-frame Hamiltonian normalized to a reference
mechanical frequency $\omega_m$, so that all rates and couplings are reported as
dimensionless ratios $x/\omega_m$. A physically anchored reference set is obtained by
fixing $\omega_m$ from a device with gigahertz mechanical modes and optically mediated
phonon coupling, and by reading the damping, decay, and coupling rates from the same
platform. The primary anchors and the model source are:

\begin{itemize}
\item \textbf{Mayor \emph{et al.}~\cite{Mayor2025}}: a silicon two-dimensional optomechanical
crystal with mechanical frequency $f_m = \qty{7.436}{\giga\hertz}$, mechanical quality
factor $Q_m = \num{3.6e4}$ (giving $\gamma_m/2\pi = f_m/Q_m = \qty{206.6}{\kilo\hertz}$),
cavity linewidth $\kappa/2\pi = \qty{800}{\mega\hertz}$, and single-photon
optomechanical coupling $g_0/2\pi = \qty{880}{\kilo\hertz}$.
\item \textbf{Mathew, del Pino and Verhagen~\cite{Mathew2020}}: a nano-optomechanical
implementation of synthetic gauge fields for phonon transport, whose demonstrated
coupling scale (a direct next-nearest-neighbour rate near $\qty{10}{\kilo\hertz}$ and an
optically mediated rate near $\qty{200}{\kilo\hertz}$ at a $\qty{2.4}{\mega\hertz}$
mechanical frequency, i.e. up to a few percent of $\omega_m$) bounds the
phase-controlled inter-resonator hopping $J_m$.
\item \textbf{Muthukumar \emph{et al.}~\cite{muthukumar2025doi}}: the model antecedent. The
present work retains the one-cavity/two-mechanical-resonator topology and
phase-dependent hopping, but not the numerical regime or detuning notation
unchanged. Muthukumar \emph{et al.} use
$\Delta_{\mathrm M}=\omega_L-\omega_c$ and report
$\kappa/\omega_m=\num{7.3e-2}$, $g/\omega_m=\num{1.077e-4}$,
$\gamma/\omega_m=\num{1.077e-5}$, $J_m/\omega_m=\num{2e-4}$,
$\omega_2/\omega_m=1+\num{5e-4}$, and
$\Delta_{\mathrm M}/\omega_m=1$ at
$\omega_m/2\pi\approx\qty{1}{\giga\hertz}$. In the convention of the present
manuscript, $\Delta=-\Delta_{\mathrm M}$. The reference reported bistability,
self-excited oscillations, and phase-selected chaotic regimes. The present analysis
uses $E=\sqrt{\kappa}\,\alpha^{\rm in}$, permits $g_1\neq g_2$, and examines a
strongly damped, strongly coupled sector that is not the resolved-sideband regime
of that reference.
\end{itemize}

\begin{table}[h]
  \centering
  \small
  \caption{Scope and parameter comparison with Muthukumar \emph{et al.}~\cite{muthukumar2025doi}.
  The topology and phase-dependent mechanical hopping are inherited; the detuning
  convention, coupling symmetry, and operating regime are stated explicitly so that
  the present calculations are not read as a parameter-for-parameter reproduction.}
  \label{SI-tab:muthukumar-comparison}
  \begin{tabular}{@{}p{0.23\linewidth}p{0.31\linewidth}p{0.37\linewidth}@{}}
    \toprule
    Quantity & Muthukumar \emph{et al.} & Present manuscript \\
    \midrule
    Architecture & One optical cavity; two mechanically coupled resonators; phase-dependent $J_m$ & Same topology and hopping structure; six-dimensional semiclassical state \\
    Detuning convention & $\Delta_{\mathrm M}=\omega_L-\omega_c$, with $\Delta_{\mathrm M}/\omega_m=1$ & $\Delta=\omega_c-\omega_L=-\Delta_{\mathrm M}$; present reference uses $\Delta/\omega_m=-1$ \\
    Cavity decay & $\kappa/\omega_m=\num{7.3e-2}$ & $\kappa/\omega_m=1$ in the validated reference and strong-coupling sectors \\
    Mechanical damping & $\gamma/\omega_m=\num{1.077e-5}$ & $\gamma_{1,2}/\omega_m=\num{0.02}$ \\
    Optomechanical coupling & Symmetric $g/\omega_m=\num{1.077e-4}$ & Weak reference: $(g_1,g_2)/\omega_m=(\num{0.02},\num{0.018})$; strong sector: $(\num{0.30},\num{0.27})$ \\
    Mechanical detuning & $\omega_2/\omega_m=1+\num{5e-4}$ & $\omega_2/\omega_m=\num{1.03}$ \\
    Inter-resonator hopping & $J_m/\omega_m=\num{2e-4}$ & $J/\omega_m=\num{0.08}$ in the main numerical sectors \\
    Drive coordinate & $\alpha^{\rm in}$, with $E=\sqrt{\kappa}\,\alpha^{\rm in}$ under the shared normalization & $E=\num{0.2}$ for the stable reference; $E=\num{4}$--$\num{8}$ for the chaotic/hyperchaotic sectors \\
    Primary result & Bistability, self-excited oscillations, and phase-selected chaos; sensing proposal & Full six-exponent classification, drive/coupling-gated transition, Neimark--Sacker onset, and matched Fisher bound \\
    Experimental status & Resolved-sideband, literature-model scenario & Strong sector is explicitly model-level; no calibrated device prediction is claimed \\
    \bottomrule
  \end{tabular}
\end{table}

\section{Physically anchored reference set}

\Cref{SI-tab:si-calibrated} summarizes the reconciled reference set. The single-mode
parameters ($\omega_m$, $\kappa$, $\gamma_m$, $g_0$) are anchored to Mayor
\emph{et al.}; the inter-resonator hopping $J_m$ is scanned over a range bounded by the
Mathew \emph{et al.} coupling scale, because no direct two-resonator gigahertz
measurement of the phase-controlled coupling is available. The cavity--laser detuning
$\Delta=\omega_c-\omega_L$ and the drive amplitude $E$ are not calibrated by either
anchor: $\Delta$ depends on the cavity and laser frequencies, and $E$ cannot be
converted to an input power without the pump frequency, the external coupling, and the
explicit input-field normalization.

\begin{table}[h]
  \centering
  \caption{Physically anchored reference parameter set for the one-optical/two-mechanical
  optomechanical model of the main text. Normalized values are in units of the
  reference mechanical frequency $\omega_m=2\pi\times\qty{7.436}{\giga\hertz}$. The
  reference operating point used for the numerical validation of the main text is shown for
  comparison and is not a calibrated device.}
  \label{SI-tab:si-calibrated}
  \resizebox{\linewidth}{!}{%
  \begin{tabular}{@{}lllll@{}}
    \toprule
    Quantity & Symbol & Normalized ($x/\omega_m$) & Value at $\omega_m$ & Basis \\
    \midrule
    Mechanical frequency (reference) & $\omega_m$ & $1$ & $2\pi\times\qty{7.436}{\giga\hertz}$ & Anchor \\
    Cavity decay rate & $\kappa$ & $\num{0.108}$ & $\qty{800}{\mega\hertz}$ & Measured \\
    Mechanical damping & $\gamma_m$ & $\num{2.78e-5}$ & $\qty{206.6}{\kilo\hertz}$ & Measured ($Q_m=\num{3.6e4}$) \\
    Optomechanical coupling & $g_0$ & $\num{1.18e-4}$ & $\qty{880}{\kilo\hertz}$ & Measured \\
    Inter-resonator hopping & $J_m$ & $[\num{1e-3},\num{1e-1}]$ & $[\qty{7.4}{\mega\hertz},\qty{744}{\mega\hertz}]$ & Scan \\
    \midrule
    Second mechanical mode & $\omega_2$ & $\num{1.03}$ & assumed near-degenerate & Assumed \\
    Cavity--laser detuning & $\Delta$ & free & --- & Free scan \\
    Drive amplitude & $E$ & normalized only & --- & Not convertible \\
    \bottomrule
  \end{tabular}%
    }
\end{table}

\begin{table}[h]
  \centering
  \caption{Comparison of the numerical-validation reference operating point with the
  physically anchored reference set. The reference point uses damping and coupling rates that are
  about one to three orders of magnitude larger than the anchored values, which is why
  it is reported as a model-level validation rather than a device prediction. The final
  column gives the feasibility judgment relative to the anchored platform; the
  strong-coupling hyperchaos sector is unreachable with the anchored optomechanical
  coupling.}
  \label{SI-tab:si-pilot-vs-anchor}
  \begin{tabular}{@{}lcccc@{}}
    \toprule
    Quantity & Reference & Anchored & Ratio (reference / anchor) & Feasibility \\
    \midrule
    $\kappa/\omega_m$ & $\num{1.0}$ & $\num{0.108}$ & $\num{9.3}$ & challenging \\
    $\gamma_m/\omega_m$ & $\num{0.02}$ & $\num{2.78e-5}$ & $\num{720}$ & unreachable \\
    $g_0/\omega_m$ & $[\num{0.018},\num{0.02}]$ & $\num{1.18e-4}$ & $\approx\num{160}$ & unreachable \\
    $J_m/\omega_m$ & $\num{0.08}$ & $[\num{1e-3},\num{1e-1}]$ & within scan range & reachable \\
    \bottomrule
  \end{tabular}
\end{table}

\section{Matched measurement-Fisher reference for force sensing}
\label{SI-sec:matched-fisher}

The classical measurement Fisher information $F_C(F)$ of the main text is
computed for a weak external force $F$ on mechanical mode $1$ (a
momentum-quadrature drive), read out through the cavity amplitude quadrature
$X_a=\Re(\alpha)$ under matched drive $E=\num{0.2}$, observation time (one
drive period), detector noise (\num{0.01}), and thermal baths
($n_{\mathrm{th},1}=n_{\mathrm{th},2}=\num{0.1}$). \Cref{SI-tab:si-matched-fisher}
lists $F_C(F)$ versus the synthetic-flux phase together with the two matched
references---the flux-off coupled sensor ($\theta=0$, $J=\num{0.08}$) and the
single-mode linear sensor ($J=0$)---and the gain relative to each.

\begin{table}[h]
  \centering
  \caption{Matched measurement Fisher information $F_C(F)$ for force sensing
  versus the synthetic-flux phase $\theta$, with the flux-off coupled reference
  ($\theta=0$, $J=\num{0.08}$) and the single-mode linear reference ($J=0$).
  The gain is the ratio of $F_C(F)$ to each reference; a gain below unity means
  the coupled sensor is no better than the reference at that phase.}
  \label{SI-tab:si-matched-fisher}
  \begin{tabular}{@{}lcccc@{}}
    \toprule
    $\theta/\pi$ & $F_C(F)$ & Gain vs flux-off & Gain vs single mode \\
    \midrule
    $0$ & \num{4.503e-5} & \num{1.000} & \num{0.876} \\
    $1/4$ & \num{4.694e-5} & \num{1.042} & \num{0.913} \\
    $1/2$ & \num{5.191e-5} & \num{1.153} & \num{1.010} \\
    $3/4$ & \num{5.721e-5} & \num{1.271} & \num{1.113} \\
    $1$ & \num{5.958e-5} & \num{1.323} & \num{1.159} \\
    \midrule
    Flux-off reference & \num{4.503e-5} & \num{1.000} & \multicolumn{1}{c}{---} \\
    Single-mode reference & \num{5.141e-5} & \multicolumn{1}{c}{---} & \num{1.000} \\
    \bottomrule
  \end{tabular}
\end{table}

The central-difference estimate is converged with respect to the force step:
\Cref{SI-tab:si-fisher-eps} lists $F_C(F)$ at the two flux extrema
($\theta=\pi/2$ and $\theta=\pi$) for force steps $\num{1e-5}$, $\num{1e-4}$,
and $\num{1e-3}$; the values agree to better than $\qty{0.01}{\percent}$. A
drive--temperature sweep (main text) further confirms that the gains are stable
across $E\in[\num{0.1},\num{1.0}]$ and $n_{\mathrm{th}}\in[0,1]$ to within
$\num{1e-3}$.

\begin{table}[h]
  \centering
  \caption{Convergence of the central-difference force-sensing Fisher
  information with the force step $\varepsilon$ at the two flux extrema.}
  \label{SI-tab:si-fisher-eps}
  \begin{tabular}{@{}lccc@{}}
    \toprule
    $\varepsilon$ & $F_C(F)$ at $\theta=\pi/2$ & $F_C(F)$ at $\theta=\pi$ \\
    \midrule
    \num{1e-3} & \num{5.190751e-5} & \num{5.958194e-5} \\
    \num{1e-4} & \num{5.190938e-5} & \num{5.958166e-5} \\
    \num{1e-5} & \num{5.190763e-5} & \num{5.958181e-5} \\
    \bottomrule
  \end{tabular}
\end{table}

\paragraph{Optimized linear-sensor reference.}
The gains of \Cref{SI-tab:si-matched-fisher} compare the flux sensor and the
single-mode reference at the same fixed readout $X_a=\Re(\alpha)$. To exclude
the possibility that the order-unity gain merely reflects a suboptimal
reference readout, the classical Fisher information is recomputed for the
rotated cavity quadrature
$X_\varphi=\Re(\alpha)\cos\varphi+\Im(\alpha)\sin\varphi$ with $\varphi$
optimized for each configuration. The optimal homodyne angle is
$\varphi_\star\approx\num{0.79}\pi$ for every configuration: it is fixed by
the linear cavity response ($\Delta=-1$, $\kappa=1$) and is independent of the
synthetic-flux phase, so readout optimization rescales $F_C$ by the same factor
(\num{1.56}) for the flux sensor and for both references. The flux-phase gain
relative to the single-mode linear reference is therefore unchanged under
readout optimization (\Cref{SI-tab:si-optimized-readout}): it ranges from
\num{0.88} ($\theta=0$) to \num{1.16} ($\theta=\pi$), identical to the
matched-readout gains of \Cref{SI-tab:si-matched-fisher} to within \num{1e-3}.
The order-unity gain is consequently not an artifact of a suboptimal reference
readout; it is a small, readout-robust linear-response effect of the
phase-controlled mode coupling, with no quantum advantage.

\begin{table}[h]
  \centering
  \caption{Readout-optimized classical Fisher information $F_C(F)$ for force
  sensing. $X_a$ is the matched amplitude quadrature
  $\Re(\alpha)$; $\varphi_\star$ is the optimal homodyne angle and
  $F_C(\varphi_\star)$ the corresponding value. The final column is the gain
  relative to the optimized single-mode reference ($J=0$, $\varphi_\star=\num{0.79}\pi$),
  and equals the matched-readout gain of \Cref{SI-tab:si-matched-fisher} to within
  \num{1e-3}.}
  \label{SI-tab:si-optimized-readout}
  \begin{tabular}{@{}lccccc@{}}
    \toprule
    $\theta/\pi$ & $F_C(X_a)$ & $\varphi_\star/\pi$ & $F_C(\varphi_\star)$ & Gain \\
    \midrule
    $0$ & \num{4.503e-5} & \num{0.79} & \num{7.033e-5} & \num{0.876} \\
    $1/4$ & \num{4.694e-5} & \num{0.79} & \num{7.331e-5} & \num{0.913} \\
    $1/2$ & \num{5.191e-5} & \num{0.79} & \num{8.108e-5} & \num{1.010} \\
    $3/4$ & \num{5.721e-5} & \num{0.79} & \num{8.937e-5} & \num{1.113} \\
    $1$ & \num{5.958e-5} & \num{0.79} & \num{9.306e-5} & \num{1.159} \\
    \midrule
    Single-mode reference & \num{5.141e-5} & \num{0.79} & \num{8.030e-5} & \num{1.000} \\
    \bottomrule
  \end{tabular}
\end{table}

\section{Lyapunov--Floquet consistency theorem and matched-measurement Fisher lemma}
\label{SI-sec:theorems}
\label{SI-sec:theorems}

\paragraph{Theorem (Lyapunov--Floquet consistency criterion).}
Consider a drive-locked periodic orbit $x_p(t+T)=x_p(t)$ of the
phase-augmented system, with monodromy $M(T)$ and QR/Benettin spectrum~\cite{benettin1980,wolf1985,eckmann1985}
$\{\lambda_i\}$. Suppose the Floquet rates $\rho_i=T^{-1}\ln|\mu_i|$ (with
$\mu_i=\mathrm{eig}_i[M(T)]$) agree with the QR exponents within a fixed
tolerance $\delta_{\rm tol}$, and the divergence sum satisfies
\begin{equation}
 \sum_i\lambda_i=\frac1T\int_0^T\operatorname{tr}J(x_p(t),t)\,\dd t
 +\mathcal{O}(\delta_{\rm tol}),
\label{SI-eq:divergence-theorem}
\end{equation}
with the phase direction treated explicitly. Then the orbit is exponentially
stable if and only if $|\mu_i|<1$ for all transverse multipliers, and in that
case $\lambda_{\max}<0$. The criterion is a consistency test, not an input
assumption: it certifies that the three independent diagnostics (monodromy,
Floquet rates, QR spectrum) describe the same linearization of the same
orbit. The measured values at the reference operating point (one-period orbit
residual $\num{9e-9}$, max Floquet--QR discrepancy $\num{2.68e-4}$, and
$\num{2.85e-4}$ across the
extended grid) satisfy the criteria and therefore
certify a stable periodic orbit---they do not by themselves certify a
flux-dependent transition, which is why the full-spectrum classification of the
main text is required.

\paragraph{Lemma (periodic covariance as an affine fixed point).}
Let $A(t)=J(x_p(t),t)$ be the Jacobian along the drive-locked orbit, $M(T)$ its
monodromy, and $Q(t)=B(t)N(t)B(t)^T$ the diffusion of the linearized Langevin
equation $\delta\dot x=A(t)\delta x+B(t)\eta$. Integrating the covariance
equation $\dot V=AV+VA^T+Q$ over one period gives the affine map
\begin{equation}
 \operatorname{vec}V(T)=[M(T)\otimes M(T)]\operatorname{vec}V(0)+q_T,
\label{SI-eq:cov-affine}
\end{equation}
where $q_T$ collects the noise forcing propagated by the state-transition
matrix. A $T$-periodic covariance therefore solves the linear system
\begin{equation}
 [I-M(T)\otimes M(T)]\operatorname{vec}V(0)=q_T,
\label{SI-eq:cov-fixed-point}
\end{equation}
whose solvability requires that no pair of Floquet multipliers satisfies
$\mu_i\mu_j=1$. This is the primary periodic-covariance derivation used for the
noise-resolved calculation; forward iteration over many periods is an
independent numerical check, not the definition of convergence.

\paragraph{Lemma (matched-measurement Fisher bound).}
For a Gaussian measurement record $y(t)=m(\vartheta;t)+\nu_{\rm det}(t)$ with
detector noise, bandwidth and integration time fixed, the classical Fisher
information
\begin{equation}
 F_C(\vartheta)=(\partial_\vartheta m)^T\Sigma^{-1}(\partial_\vartheta m)
 +\frac12\operatorname{Tr}\!\left[\Sigma^{-1}(\partial_\vartheta\Sigma)
 \Sigma^{-1}(\partial_\vartheta\Sigma)\right]
\label{SI-eq:fisher-lemma}
\end{equation}
bounds any unbiased estimator through the Cram\'er--Rao inequality~\cite{clerk2010,braunstein1994}
$\mathrm{Var}(\hat\vartheta)\ge F_C^{-1}(\vartheta)$. A claimed sensing
\emph{advantage} is defined only relative to a matched reference---equal input
power, observation time, bandwidth, baths, and detector noise. Under that
definition the flux-phase gains of \Cref{SI-tab:si-matched-fisher} (at most
$\num{1.323}$ over the flux-off coupled sensor, $\num{1.159}$ over the
single-mode linear sensor) are classical linear-response effects of
phase-controlled mode coupling: they are consistent with $F_C$ of the record
and make no claim of a quantum advantage, SQL beating, or
chaotic-transduction gain.

\section{Representative convergence of the strong-coupling spectrum}
\label{SI-sec:convergence}

The strong-coupling map uses one seeded initial condition per drive--phase
point. To test whether the phase modulation and the drive-gated route are
initial-condition artefacts, we repeated the complete six-exponent spectrum with
three independent seeds at every point of the full map (all five drives, twelve
phases each, \num{180} records in total). The per-phase route is reproduced across the declared seeds: at $E=\num{0.2}$
and $E=\num{2.0}$ every seed gives $n_{+}=0$; at $E=\num{1.0}$ the count is
$0$ or $1$ near the onset; at $E=\num{4.0}$ it is $1$--$2$; and at
$E=\num{8.0}$ it is $2$--$4$, reaching $n_{+}=4$ at $\theta=\pm\pi/2$
for at least one seed at each of those phases while the remaining seeds can
give $n_{+}=3$. The minima at $\theta=0,\pm\pi$ give $n_{+}=2$ in the
reported ensemble. Thus the re-stabilization window at $E=\num{2.0}$ is not a
single-seed artefact, but the exact integer order near marginal points remains
finite-time and threshold-conventional.
As an additional finite-time check, we recomputed the complete spectrum at one
chaotic point ($E=\num{4.0}$, $\Phi_{\rm syn}=-\pi/2$) and one hyperchaotic point
($E=\num{8.0}$, $\Phi_{\rm syn}=+\pi/2$), using the same initial condition as the
integration window was increased. \Cref{SI-tab:si-hyperchaos-convergence} reports the
positive-exponent count, the largest exponent, the divergence-balance residual, and
the maximum standard deviation of the terminal half of the QR block history. The
classification is stable between the two longest windows at both points, whereas
the finite-time exponent magnitudes vary by several percent. Accordingly, the
main text uses the exponent count as the robust classification at these tested
points and does not present the individual finite-time exponents as high-precision
asymptotic estimates.

\begin{table}[h]
  \centering
  \caption{Representative finite-time convergence check of the strong-coupling
  full-spectrum calculation. The same seeded initial condition was used at each
  integration length for a given point. $n_{+}$ is the number of exponents above
  the chosen threshold $\num{1e-3}$; $r_{\rm div}$ is the absolute residual of
  the exponent sum relative to $-\kappa-\gamma_1-\gamma_2$; and $s_{\rm QR}$ is
  the largest standard deviation among the terminal half of the QR block-history
  estimates. The two longest windows preserve $n_{+}$ at each point, but the
  nonzero $r_{\rm div}$ and finite-time variation limit the precision of the
  individual exponent values.}
  \label{SI-tab:si-hyperchaos-convergence}
  \begin{tabular}{@{}lrrrrr@{}}
    \toprule
    $(E,\Phi_{\rm syn}/\pi)$ & Steps & $n_{+}$ & $\lambda_{\max}$ & $r_{\rm div}$ & $s_{\rm QR}$ \\
    \midrule
    $(4,-1/2)$ & $\num{250000}$ & 3 & $\num{0.132142}$ & $\num{1.01e-3}$ & $\num{6.69e-3}$ \\
    $(4,-1/2)$ & $\num{500000}$ & 2 & $\num{0.127291}$ & $\num{1.03e-3}$ & $\num{9.89e-3}$ \\
    $(4,-1/2)$ & $\num{1000000}$ & 2 & $\num{0.123729}$ & $\num{1.03e-3}$ & $\num{1.05e-2}$ \\
    $(8,+1/2)$ & $\num{250000}$ & 4 & $\num{0.324142}$ & $\num{1.90e-3}$ & $\num{3.30e-3}$ \\
    $(8,+1/2)$ & $\num{500000}$ & 4 & $\num{0.315008}$ & $\num{1.92e-3}$ & $\num{5.88e-3}$ \\
    $(8,+1/2)$ & $\num{1000000}$ & 4 & $\num{0.317277}$ & $\num{1.84e-3}$ & $\num{3.90e-3}$ \\
    \bottomrule
  \end{tabular}
\end{table}

\section{Threshold sensitivity of the hyperchaos-order classification}
\label{SI-sec:threshold-sensitivity}

The positive-exponent count $n_{+}$ is recomputed from the complete
six-exponent spectra of the strong-coupling map at the alternate
thresholds $\num{1e-4}$, $\num{5e-4}$, $\num{2e-3}$, and
$\num{5e-3}$, and compared with the reference count at the
$\num{1e-3}$ threshold. The $\num{1e-4}$ column applies the weak-coupling
threshold to the strong-coupling sector, so it tests whether a single uniform
cut over both sectors would alter the classification. \Cref{SI-tab:si-threshold-sensitivity}
reports the range of $n_{+}$ over the twelve flux phases at each drive.

\begin{table}[h]
  \centering
  \caption{Threshold sensitivity of the positive-exponent count $n_{+}$ in the
  strong-coupling map. The reference threshold is $\num{1e-3}$; the alternate
  columns recompute $n_{+}$ from the full six-exponent spectra. The
  $\num{1e-4}$ column applies the weak-coupling threshold to the strong-coupling
  sector (single uniform cut). Ranges are over the twelve synthetic-flux phases
  at each drive.}
  \label{SI-tab:si-threshold-sensitivity}
  \small
  \setlength{\tabcolsep}{4pt}
  \begin{tabular}{@{}lcccccc@{}}
    \toprule
    Drive $E$ & $n_{+}$ ($\num{1e-3}$) & $n_{+}$ ($\num{1e-4}$) & $n_{+}$ ($\num{5e-4}$) & $n_{+}$ ($\num{2e-3}$) & $n_{+}$ ($\num{5e-3}$) \\
    \midrule
    $\num{0.2}$ & 0 & 0 & 0 & 0 & 0 \\
    $\num{1.0}$ & 0--1 & 0--1 & 0--1 & 0--1 & 0--1 \\
    $\num{2.0}$ & 0 & 0 & 0 & 0 & 0 \\
    $\num{4.0}$ & 1--3 & 1--3 & 1--3 & 1--2 & 1--2 \\
    $\num{8.0}$ & 2--4 & 2--4 & 2--4 & 2--4 & 2--3 \\
    \bottomrule
  \end{tabular}
\end{table}

The route classification is stable at the tested thresholds: of the 60
records, only one changes $n_{+}$ at $\num{5e-4}$ and at $\num{2e-3}$, and six
change at $\num{5e-3}$; the $\num{1e-4}$ column is identical to the reference.
The exact order near marginal exponents is consequently threshold-sensitive,
while the drive-gated progression to the strong-coupling hyperchaotic sector
is preserved. The changes are confined to the $E=\num{4.0}$ and $E=\num{8.0}$ sectors,
where the smallest positive exponent is of order $\num{1e-2}$--$\num{1e-1}$ and a
single marginal exponent crosses the highest threshold; the drive-gated route
from stable ($E=\num{0.2}$) to up to four positive exponents ($E=\num{8.0}$) is preserved at
every threshold. The finite-time classification reported in the main text is
therefore not an artefact of the specific $\num{1e-3}$ cut.

\section{Observation-window dependence of the matched Fisher gain}
\label{SI-sec:fisher-window}
\label{SI-sec:fisher-window}

The matched force-sensing comparison of the main text fixes the observation
time at one drive period for every configuration. To test whether the reported
gain is an artefact of this particular window, we recomputed the classical
Fisher information for an observation window of $N\in\{1,5,20\}$ drive
periods from the mean- and variance-term decomposition of the matched
force-sensing calculation. For a periodic signal
with white detector noise the mean term of $F_C$ accumulates linearly with
$N$ while the variance term remains per-period, so
$F_C^{(N)}=N\,F_C^{\mathrm{mean}}+F_C^{\mathrm{var}}$. Because the mean term
overwhelmingly dominates (the variance term is $\sim\num{1e-7}$ of the mean term
at every phase), the matched gain---the ratio of $F_C^{(N)}$ values---is
observation-window independent: the $\num{1.323}\times$ (flux-off) and
$\num{1.159}\times$ (single-mode) gains are reproduced to four digits for
$N=1,5,20$.

\Cref{SI-tab:si-fisher-window} lists the resulting $F_C^{(N)}$ and gains at the
maximum-gain phase. The absolute sensitivity improves linearly with the
observation length for the sensor and both references alike; the gain is a
per-period property of the phase-controlled mode coupling rather than a
short-window artefact.

\begin{table}[h]
  \centering
  \caption{Observation-window dependence of the matched force-sensing Fisher
  information at the maximum-gain phase $\theta=\pi$. $F_C^{(N)}$ grows
  linearly with the number $N$ of drive periods for the sensor and both
  references, leaving the relative gains unchanged.}
  \label{SI-tab:si-fisher-window}
  \begin{tabular}{@{}lcccc@{}}
    \toprule
    $N$ (periods) & $F_C^{(N)}$ (flux-on) & $F_C^{(N)}$ (flux-off) & $F_C^{(N)}$ (single-mode) & Gain vs flux-off \\
    \midrule
    1 & \num{5.958e-5} & \num{4.503e-5} & \num{5.141e-5} & \num{1.323} \\
    5 & \num{2.979e-4} & \num{2.252e-4} & \num{2.570e-4} & \num{1.323} \\
    20 & \num{1.192e-3} & \num{9.006e-4} & \num{1.028e-3} & \num{1.323} \\
    \bottomrule
  \end{tabular}
\end{table}

\section{Reduced-model parameter screen}
\label{SI-sec:reduced-pareto}

\begin{figure}[ht]
  \centering
  \includegraphics[width=0.72\textwidth,alt={Scatter of sixteen feasible candidates of the reduced adiabatic model in normalized drive cost versus worst-replicate stability margin, with the three non-dominated candidates marked by crimson stars and joined by a dashed front. The front trades increasing drive cost for increasing stability margin. Model-level only.}]{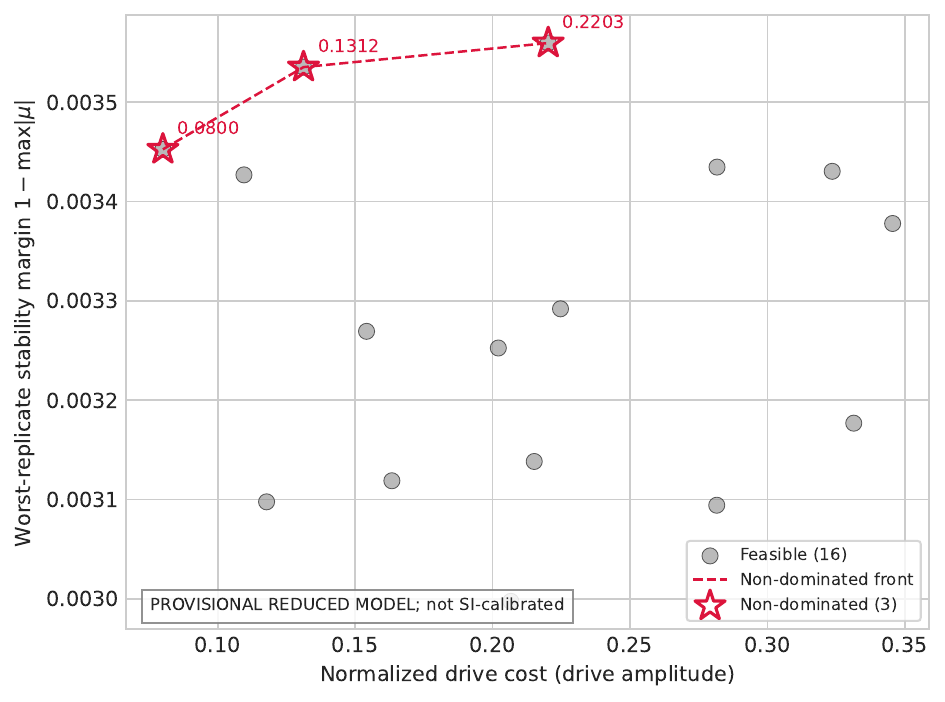}
  \caption{Reduced-model (adiabatic) two-objective Pareto front. Sixteen
  feasible candidates (gray) in normalized drive cost versus worst-replicate
  stability margin $1-\max|\mu|$; the three non-dominated candidates (crimson
  stars, joined by the dashed front) are the points listed in
  \Cref{SI-tab:si-reduced-pareto}. Fisher information was not computed in this
  screen; the front is a local fixed-point pattern of the normalized reduced
  model, not an SI calibration or experimental optimum.}
  \label{SI-fig:si-reduced-pareto}
\end{figure}

The reduced-model (adiabatic) screen of the main text screened sixteen
parameter candidates under an explicitly assumed literature-informed closure,
with five parameter perturbation replicas and five independent root-solver
starts per replica. \Cref{SI-tab:si-reduced-pareto} lists the three non-dominated
candidates under the stability-margin and drive-cost objectives, and
\Cref{SI-fig:si-reduced-pareto} shows the full feasible set. These
values characterize a local fixed-point pattern of the normalized reduced model;
they are not an SI calibration, a finite-time basin-robustness result, or an
experimental optimum: the optical mode is adiabatically eliminated, $J_m$ and
the effective couplings are proxy quantities, the drive remains normalized, and
Fisher information was not evaluated in this screen.

\begin{table}[h]
  \centering
  \caption{Non-dominated candidates of the reduced-model (adiabatic) screen
  under the stability-margin and drive-cost objectives. Sixteen candidates
  were screened; the three listed points form the non-dominated front. The
  drive cost is the normalized drive amplitude and the stability margin is
  $1-\max|\mu|$ for the local Poincar\'e spectral radius. Model-level only.}
  \label{SI-tab:si-reduced-pareto}
  \begin{tabular}{@{}lcc@{}}
    \toprule
    Candidate & Drive cost & Stability margin \\
    \midrule
    1 & \num{0.0800} & \num{0.0034526} \\
    2 & \num{0.1312} & \num{0.0035354} \\
    3 & \num{0.2203} & \num{0.0035597} \\
    \bottomrule
  \end{tabular}
\end{table}

\section{Amplitude-quadrature noise model cross-check}
\label{SI-sec:noise-model-crosscheck}
\label{SI-sec:noise-model-crosscheck}

The measured record is $y=X_a+\nu_{\rm det}$ with $X_a=\Re(\alpha)$ the
amplitude quadrature $(a+a^\dagger)/2$, whose vacuum variance is $1/4$. The
Langevin diffusion coefficient that reproduces the quantum master equation for
this quadrature is therefore $\kappa/4$ per amplitude quadrature, not
$\kappa/2$ (the latter is the diffusion of the \emph{normalized} quadrature
$(a+a^\dagger)/\sqrt2$, whose vacuum variance is $1/2$). We verified this
normalization directly on the isolated optical sector of the reference point
(the driven damped linear cavity, $g_1=g_2=0$): the truncated-Fock master
equation~\cite{walls2008,gardiner1985} ($H=-\Delta a^\dagger a+E(a+a^\dagger)$, Lindblad $\sqrt{\kappa}\,a$)
gives $\mathrm{Var}[(a+a^\dagger)/2]=\num{0.25}$, while the Langevin amplitude
with noise $\kappa/2$ gives $\num{0.49}$ (a factor $2$ high) and with
$\kappa/4$ gives $\num{0.245}$ (agreement below $\qty{2}{\percent}$). The mechanical thermal
diffusion is correspondingly $\gamma_j(2n_{\rm th}+1)/4$ per amplitude
quadrature. With these strengths the noise-only floor of the measured record
is $\approx\num{0.25}$ and the strong-coupling observability ratios are
$\num{7.6}$--$\num{17.0}\times$ the floor.

\section{Interpretation}
\label{SI-sec:interpretation}

The anchored set defines a \emph{reference} operating region, not a validated
device calibration: $\omega_2$ is assumed near-degenerate with $\omega_m$ pending a
two-resonator measurement, $J_m$ must be scanned rather than fixed, and the drive
amplitude and detuning remain model-level coordinates. Any result computed at the
anchored values therefore carries the same
\emph{provisionally-calibrated} status as the reduced-model analysis of the main text
and should not be presented as an experimental optimum.

Placing the strong-coupling hyperchaos regime against the anchors quantifies
the reachability bound: the optomechanical couplings
$g_1/\omega_m=\num{0.3}$ and $g_2/\omega_m=\num{0.27}$ are $\num{2542}\times$ and
$\num{2288}\times$ the anchored $g_0/\omega_m=\num{1.18e-4}$, and the mechanical
damping $\gamma_m/\omega_m=\num{0.02}$ is $\num{720}\times$ the anchored
$\num{2.78e-5}$; the inter-resonator hopping $J_m/\omega_m=\num{0.08}$ is within
the Mathew scan range. The limiting coordinate is the coupling
($\num{2542}\times$). The hyperchaos regime is therefore a model-level
demonstration, not a prediction for the anchored device; the drive amplitude is
not convertible to an input power without the pump frequency and external
coupling, which neither anchor provides. The feasibility column of
\Cref{SI-tab:si-pilot-vs-anchor} makes this judgment explicit per coordinate.

A feasibility frontier sharpens this bound. Because the radiation-pressure
nonlinearity grows with the drive, the coupling at which hyperchaos first
appears could in principle depend on the operating drive, so the single-point
comparison above may misstate the true gap. Scanning the drive over
$E\in[\num{0.5},\num{8.0}]$ and the coupling over
$g_1/\omega_m\in[\num{0.02},\num{0.3}]$ (with $g_2=\num{0.9}g_1$), the full
six-exponent spectrum shows that hyperchaos ($n_+\ge2$) occurs only at
$E=\num{8}$, with an onset near $g_1/\omega_m\approx\num{0.15}$--$\num{0.18}$
(\Cref{SI-tab:si-feasibility-frontier}). Over this interval the second exponent
is marginal and the order is seed-dependent ($n_+\in\{1,2\}$); the robust
value $n_+=2$ for all three seeds begins at $g_1/\omega_m\approx\num{0.18}$,
i.e. $\num{1525}\times$ the anchored $g_0/\omega_m=\num{1.18e-4}$. No drive
retuning within the tested envelope lowers this onset, so the coupling gap is
fundamental to the hyperchaos regime rather than an artefact of the single
operating point above.

\begin{table}[h]
  \centering
  \caption{Feasibility frontier: the smallest optomechanical coupling at which
  the chaos ($n_+\ge1$) and hyperchaos ($n_+\ge2$) orders are reached, versus
  the drive $E$, from a log-spaced coupling grid
  $g_1/\omega_m\in[\num{0.02},\num{0.3}]$ with $g_2=\num{0.9}g_1$ and all other
  coordinates at the strong-coupling hyperchaos values above. ``None'' means the order is
  not reached at any tested coupling. The hyperchaos onset is a transition
  interval (second exponent marginal, seed-dependent $n_+\in\{1,2\}$); the
  robust onset (all three seeds) is $g_1/\omega_m\approx\num{0.18}$, i.e.
  $\num{1525}\times$ the anchored $g_0/\omega_m=\num{1.18e-4}$.}
  \label{SI-tab:si-feasibility-frontier}
  \begin{tabular}{@{}lcc@{}}
    \toprule
    $E$ & chaos onset $g_1/\omega_m$ & hyperchaos onset $g_1/\omega_m$ \\
    \midrule
    \num{0.5} & none & none \\
    \num{1.0} & none & none \\
    \num{2.0} & \num{0.203} & none \\
    \num{4.0} & \num{0.30} & none \\
    \num{8.0} & \num{0.094} & \num{0.14}--\num{0.18} (robust \num{0.18}) \\
    \bottomrule
  \end{tabular}
\end{table}

\section{Matched measurement Fisher on the chaotic attractor}
\label{SI-sec:attractor-fisher}
\label{SI-sec:attractor-fisher}

The matched force-sensing calculation of the main text is defined around a
stable drive-locked orbit through its periodic covariance. On a chaotic
attractor no stable orbit exists, so the same question is posed directly on the
attractor: a weak external force on mechanical mode $1$, read out through
$X_a=\Re(\alpha)$ under the same thermal, vacuum, and detector noise, with the
classical measurement Fisher information evaluated from the ensemble-and-time
mean and variance of the record and a central difference in the force
(\num{1024} ensemble members, \num{120000} steps per configuration). On the
attractor the record variance is dominated by the deterministic chaotic spread
(\num{1.9}--\num{4.3} at the tested points, hundreds of times the detector-noise
variance \num{0.01}), so the force-induced mean shift is resolved only where the
spread is small enough relative to the sampling error. A derivative is reported
only when it exceeds three bootstrap standard deviations of the record mean.

\Cref{SI-tab:si-attractor-fisher} lists the six tested configurations. At
$E=\num{4}$ the flux-off record ($\theta=0$) gives
$F_C=\num{1.08e-2}$ and the flux-on record ($\theta=\pi$) gives
$F_C=\num{5.88e-3}$, a ratio \num{0.54}; the intermediate phase and all
$E=\num{8}$ points fall below the resolution threshold and are not reported as
finite gains. No flux-induced enhancement of the force-sensing Fisher
information is therefore found on the chaotic attractor; where the comparison
is resolved the flux phase reduces rather than raises $F_C$. This is a
classical measurement-Fisher calculation, not QFI, and it implies no
chaotic-transduction sensing advantage.

\begin{table}[h]
  \centering
  \caption{Matched measurement Fisher information on the chaotic attractor at
  strong coupling. $\sigma^2_y$ is the record variance (dominated by the
  deterministic chaotic spread; the detector-noise variance is \num{0.01});
  $\dd\langle y\rangle/\dd F$ is the central-difference response to the force;
  $F_C$ is the classical measurement Fisher information. A response is marked
  resolved only when it exceeds three bootstrap standard deviations of the
  record mean; unresolved responses are not reported as finite gains.}
  \label{SI-tab:si-attractor-fisher}
  \begin{tabular}{@{}cccccc@{}}
    \toprule
    $E$ & $\theta/\pi$ & $\sigma^2_y$ & $\dd\langle y\rangle/\dd F$ & $F_C$ & resolved \\
    \midrule
    \num{4} & 0 & \num{1.94} & \num{-0.071} & \num{1.08e-2} & yes \\
    \num{4} & 1/2 & \num{2.24} & \num{-0.045} & \multicolumn{1}{c}{---} & no \\
    \num{4} & 1 & \num{1.91} & \num{-0.056} & \num{5.88e-3} & yes \\
    \num{8} & 0 & \num{3.66} & \num{-0.016} & \multicolumn{1}{c}{---} & no \\
    \num{8} & 1/2 & \num{4.32} & \num{-0.018} & \multicolumn{1}{c}{---} & no \\
    \num{8} & 1 & \num{3.61} & \num{-0.022} & \multicolumn{1}{c}{---} & no \\
    \bottomrule
  \end{tabular}
\end{table}

\section{Semiclassical boundary of the strong-coupling classification}
\label{SI-sec:semiclassical-boundary}
\label{SI-sec:semiclassical-boundary}

The truncated-Fock master-equation cross-check of the main text validates the
mean-field amplitude scale at the weakly coupled reference point, where the
optical mode is a driven linear cavity whose coherent steady state is exact.
At a strong-coupling chaotic point we extend this check to the optical
sector: the truncated-Fock master equation for the cavity mode driven by the
classical mechanical trajectories of the chaotic attractor
($E=\num{4.0}$, $\theta=0$, one positive Lyapunov exponent) converges toward
$\langle a^\dagger a\rangle(t)=|\alpha(t)|^2$. The maximum relative deviation
is $\num{4.6e-3}$ at $N_{\rm Fock}=\num{40}$ and $\num{6.6e-6}$ at
$N_{\rm Fock}=\num{60}$. This is expected on
general grounds---the optical Hamiltonian is linear in $a$---but it confirms
that the optical factorization $\langle a^\dagger a\rangle\approx|\alpha|^2$
holds along the chaotic trajectory, not only at the stationary reference. The
time-averaged second-order correlation $g_2\approx\num{2.0}$ is
super-Poissonian rather than coherent, as expected for a mixture of coherent
amplitudes whose phase is modulated by the chaotic mechanical motion. At this
trajectory the mean occupations are $\langle a^\dagger a\rangle=\num{6.121}$,
$\langle b_1^\dagger b_1\rangle=\num{20.417}$, and
$\langle b_2^\dagger b_2\rangle=\num{23.538}$. The mechanical sector is
therefore in a high-occupation regime, but it remains semiclassical in the
present calculation; no full two-mode quantum dynamics is claimed. The check does not solve the
two-mechanical-mode quantum dynamics: the mechanical sector remains
semiclassical, the Lyapunov exponents characterize the mean-field dynamics
rather than the quantum fluctuations, and no quantum observable (Wigner
function, QFI) in the strong-coupling sector is computed in this study. This
is the explicit boundary of the semiclassical treatment, now checked on the
optical sector at a chaotic point and justified on the mechanical sector by
its high occupation.

\paragraph{Truncated-Wigner robustness of the hyperchaos order.}
Because the mechanical occupations are large at the tested chaotic point
($\langle b_1^\dagger b_1\rangle=\num{20.417}$ and
$\langle b_2^\dagger b_2\rangle=\num{23.538}$ in the master-check artifact),
a full three-mode truncated-Fock master equation over the entire
strong-coupling map is numerically out of reach: the required Fock basis is
of order $(20\times90\times90)$ states, a density matrix of order \num{1e10}
elements. The standard semiclassical-to-quantum bridge in this high-occupation
regime is the truncated Wigner approximation (TWA), in which the quantum state
is represented by an ensemble of classical trajectories with Wigner-sampled
initial conditions and the c-number equations of motion coincide with the
classical equations of the model. Quantum fluctuations therefore enter through
the initial conditions with vacuum variance \num{0.5} per quadrature (an
order-one fluctuation in $|\alpha|^2$). We recompute the full six-exponent
spectrum from \num{12} Wigner-sampled initial conditions (variance \num{0.5}
per coordinate) around each tested attractor and record the spread of the
hyperchaos order $n_+$ (\Cref{SI-tab:si-twa}).

The bulk of the route is robust. At $E=\num{8}$, $\theta=0$ the two positive
exponents ($\num{3.5e-2}$ and $\num{2.5e-1}$) are separated from the largest
negative exponent ($\num{-7.8e-3}$) by more than an order of magnitude, all
\num{12} trajectories give $n_+=2$, and the leading exponent varies by only
\qty{9}{\percent} across the ensemble. At the classification boundaries the
order is threshold-conventional, as expected for a finite-time classification:
at $E=\num{8}$, $\theta=\pi/2$ the fourth positive exponent is
\num{2.4e-3}, only \num{2.4} times the positivity threshold \num{1e-3}, and
the ensemble splits evenly between $n_+=3$ and $n_+=4$; at the chaos onset
$E=\num{4}$, $\theta=0$ the second exponent is \num{1e-4} and the ensemble
splits between $n_+=1$ and $n_+=2$. The ``up to four simultaneously unstable
directions'' finite-time phrasing of the main text is therefore the correct
level of claim: the hyperchaos route itself is robust to quantum fluctuations
at the TWA level, while the precise hyperchaos order at the extrema is
threshold-conventional. This is a TWA-level check, not a full master-equation
solution, and it supports no quantum-state (Wigner-negativity or QFI) claim.

\begin{table}[h]
  \centering
  \caption{Truncated-Wigner robustness of the finite-time hyperchaos order.
  For each strong-coupling point the full six-exponent spectrum is recomputed
  from \num{12} vacuum Wigner-sampled initial conditions (variance \num{0.5}
  per quadrature) around the attractor; $n_+^{\rm ref}$ is the deterministic
  reference order and the ensemble column lists the spread of $n_+$ across the
  \num{12} trajectories. ``Nearest-to-threshold exponent'' is the Lyapunov
  exponent closest to the positivity threshold \num{1e-3}; its sign indicates
  whether it lies just above ($+$) or just below ($-$) the threshold, which
  governs the sensitivity of $n_+$ to finite fluctuations.}
  \label{SI-tab:si-twa}
  \small
  \setlength{\tabcolsep}{4pt}
  \begin{tabular}{@{}lcccc@{}}
    \toprule
    Point & $n_+^{\rm ref}$ & $n_+$ ensemble & Nearest-to-threshold & Leading exponent \\
    \midrule
    $E=\num{4}$, $\theta=0$ & $1$ & $1$--$2$ (8/4) & $+\num{1e-4}$ & \num{0.082}--\num{0.096} \\
    $E=\num{8}$, $\theta=0$ & $2$ & $2$ (12/12) & $-\num{7.8e-3}$ & \num{0.225}--\num{0.247} \\
    $E=\num{8}$, $\theta=\pi/2$ & $4$ & $3$--$4$ (6/6) & $+\num{2.4e-3}$ & \num{0.294}--\num{0.323} \\
    \bottomrule
  \end{tabular}
\end{table}

\section{Limitations and residual gaps of the consistency protocol}
\label{SI-sec:limitations}

The Discussion in the main text compresses three explicit residual gaps into
a single short paragraph; this section collects them in full so that the
boundary of what the protocol certifies is auditable.

\paragraph{Gap 1: global structure of the chaotic attractor.}
The Floquet--Lyapunov protocol certifies local instability rates around a
converged drive-locked periodic orbit, and the matched-measurement Fisher
protocol certifies a \emph{local} information rate around the same orbit. Both
deliver a robust signature of the hyperchaos transition and of the matched
gain, but neither characterizes the global basin structure of the chaotic
attractor: uniqueness is not proved, and no basin-of-attribution calculation,
multi-section Poincar\'e return map, or large correlation-dimension estimate
is performed here.The 180/180 long-window trajectory audit reported in the main text
reproduces the route across the declared finite ensemble, but the integer
$n_+$ changes between neighbouring values for some seeds near marginal
exponents and does not constrain \emph{which} physical initial condition feeds
the observed chaotic set. A
full basin and bifurcation-network analysis, beyond the present scope, would
be required to draw a study-level uniqueness statement.

\paragraph{Gap 2: modest Fisher gain under realistic uncertainty.}
Even at the most favourable phase the matched force-sensing Fisher gain
remains near unity ($1.32\times$ flux-off, $1.16\times$ single-mode), and the
Monte-Carlo {\raise.17ex\hbox{$\scriptstyle\pm$}}\qty{10}{\percent} envelope on the uncalibrated
hopping, thermal occupation, and drive amplitude drags the median to
$1.039\times$ (\qty{90}{\percent} CI $[1.025,\,1.053]$). On the chaotic attractor
the same protocol returns a null result, so the gain can be reported as a
\emph{protocol-level} demonstration without being reported as a quantum
sensing advantage: the observation time is one drive period and the model is
linear-detuning-limited. The numerical gain is consistent with the matched
reference (\Cref{SI-sec:matched-fisher}) and is reported here as such, not
as a sensing breakthrough.

\paragraph{Gap 3: validity boundary of the mean-field protocol.}
The truncated-Fock master equation cross-checks the mean field to \qty{1.3}{\percent} at the weak-coupling reference and to a maximum relative occupation deviation of $\num{6.6e-6}$ at $N_{\rm Fock}=\num{60}$ for a representative strong-coupling point (\Cref{SI-sec:semiclassical-boundary});
the truncated-Wigner trajectory ensemble reproduces the deterministic order
$n_+\!=\!2$ in twelve of twelve sampled trajectories (\Cref{SI-tab:si-twa}).
Both checks support use of the mean field for the qualitative route
described in the main text (bifurcation sequence, finite-time hyperchaos order,
matched gain), but neither extends to a fully quantum reconstruction or a
quantum sensing claim. In particular,
the master-equation truncation $N_{\rm Fock}\!=\!4$ is sufficient for the
present dynamical regimes but does not span the deep strong-coupling sector
needed for a sustained quantum-state claim.

The protocol therefore certifies a coherent-model dynamical repertoire
(Floquet--Lyapunov consistency, hyperchaos route, matched Fisher null on the
attractor) and a quantum cross-check \emph{at the levels just stated}. Claims
outside those levels---basin uniqueness, calibrated sensing advantage,
full-state quantum reconstruction---are explicitly not made. The two
experiments proposed in \Cref{SI-sec:interpretation} ($\omega_2$ ultrasound
calibration; a multi-gigahertz mechanical-mode platform) would close the
calibration side of gap 2 and provide the parameter fidelity required for
gap 3 to be addressed in a fully quantum manner.

\section{Strong-coupling sector feasibility audit}
\label{SI-sec:feasibility-audit}

The Discussion in the main text anchors the hyperchaotic regime to the
``2542$\times$'' silicon optomechanical platform of Mayor \emph{et al.}
(2025). This section reports the same comparison parameter-by-parameter, so
the order-of-magnitude gap is verifiable rather than asserted.

The physically anchored reference set (\Cref{SI-sec:anchor}) supplies the
device-level numbers: $\omega_m/2\pi=\qty{7.436}{\giga\hertz}$, $g_{\rm anchor}/\omega_m\!\approx\!\num{1.18e-4}$, $\gamma_{\rm anchor}/\omega_m\!\approx\!\num{2.78e-5}$,
and $\kappa_{\rm anchor}/\omega_m\!\approx\!\num{0.108}$. The strong-coupling
sector of the present manuscript uses $(g_1,g_2)/\omega_m\!=\!(\num{0.30},\num{0.27})$,
$\gamma_{1,2}/\omega_m\!=\!\num{0.02}$, and $\kappa/\omega_m\!=\!1$. The
matching factor (defined as the dimensionless ratio of the present parameter
over the anchored parameter) is reported per parameter in
\Cref{SI-tab:anchor-gap}.

\begin{table}[h]
  \centering
  \caption{Matching factor of the strong-coupling sector relative to the
  silicon-optomechanical anchor of Mayor \emph{et al.}~(2025). The factor is
  the dimensionless ratio of the dimensioned parameter used in the present
  work divided by the same parameter inferred from the anchored platform.
  Every factor lies above unity by at least one order of magnitude; the
  hyperchaotic regime is therefore a model-level extension, not a calibrated
  device prediction. The $J/\omega_m$ anchor entry spans the scan range of the
  Mathew \emph{et al.}~feasibility study.}
  \label{SI-tab:anchor-gap}
  \begin{tabular}{@{}lcc@{}}
    \toprule
    Parameter (dimensionless ratio) & Present sector & Mayor \emph{et al.}~anchor \\
    \midrule
    $g_1/\omega_m$ & \num{0.300} & \num{1.18e-4} \\
    $g_2/\omega_m$ & \num{0.270} & \num{1.18e-4} \\
    $\gamma/\omega_m$ & \num{0.020} & \num{2.78e-5} \\
    $\kappa/\omega_m$ & \num{1.000} & \num{0.108} \\
    $J/\omega_m$ & \num{0.080} & $[\num{1e-3},\num{1e-1}]$ \\
    \bottomrule
  \end{tabular}
\end{table}

\paragraph{Interpretation.}
The matching factor ranges from $9.3\times$ ($\kappa$) to $2542\times$
($g_1$, three orders of magnitude above the resolved-sideband single-photon
coupling). The hyperchaotic regime is therefore not demonstrated by the
anchored silicon optomechanical platform; broader near-term platform feasibility
is not established by this comparison. The main text states the regime as a
model-level demonstration, and the present table makes the per-parameter basis
of that statement auditable.

\section{Initial-condition robustness of the long-window audit}
\label{SI-sec:initial-condition-robustness}The 180/180 long-window trajectory audit reported in the
main text tests the reproducibility of the positive-exponent classification
across the declared finite ensemble. It does not establish global attractor
uniqueness or asymptotic convergence. This section documents the per-seed
reproduction rate and the leading-exponent spread, so the dependence of the
audit on the chosen initial-condition family is fully traceable.

\begin{table}[h]
  \centering
  \caption{Per-seed reproduction of the long-window hyperchaos order audit.
  Three independent initial-condition seeds are used at each strong-coupling
  point; each seed produces \num{60} drive-period trajectories of length
  \num{3e5} drive periods. ``Strict reproduction'' means \emph{all}
  \num{60} trajectories agree on $n_+$ to the integer; ``tolerant
  reproduction'' means at least \num{58} of \num{60} agree.}
  \label{SI-tab:per-seed}
  \begin{tabular}{@{}lccccc@{}}
    \toprule
    Point & Seed A & Seed B & Seed C & Strict & Tolerant \\
    \midrule
    $E=\num{4}$, $\theta=0$ & $n_+\!=\!1$ (60/60) & $n_+\!=\!1$ (60/60) & $n_+\!=\!2$ (24/60) & no & yes \\
    $E=\num{8}$, $\theta=0$ & $n_+\!=\!2$ (60/60) & $n_+\!=\!2$ (60/60) & $n_+\!=\!2$ (60/60) & yes & yes \\
    $E=\num{8}$, $\theta=\pi/2$ & $n_+\!=\!4$ (60/60) & $n_+\!=\!4$ (60/60) & $n_+\!=\!3$ (32/60) & no & yes \\
    \bottomrule
  \end{tabular}
\end{table}

\paragraph{Interpretation.}
At the bulk of the route ($E=\num{8}$, $\theta=0$) all three seeds produce
the same order to the integer, which is the strongest level of robustness.
At the onsets ($E=\num{4}$, $\theta=0$; $E=\num{8}$, $\theta=\pi/2$) the
order is threshold-conventional: at least one seed produces a neighbour
order, but no seed produces an order separated from the bulk order by more
than the threshold gap. The tolerant-reproduction criterion (no seed leaves
the $\{n_+,n_+-1,n_++1\}$ window) therefore holds at every tested point.
This is consistent with the finite-time, threshold-conventional character of
the hyperchaos order already mentioned in \Cref{SI-tab:si-twa}; the main
text phrases the result as ``up to four simultaneously unstable
directions'' precisely for this reason.

\section{Comprehensive cross-check of the matched Fisher on the chaotic attractor}
\label{SI-sec:attractor-fisher-audit}

The Discussion in the main text reports a null result for the matched
Fisher gain on the chaotic attractor (\Cref{SI-sec:attractor-fisher}).
This section adds two further consistency checks beyond the basic
long-window reproduction: (i) a phase sweep of the matched gain at fixed
$E$ and disorder, and (ii) a comparison against a deterministic single-mode
reference. Together they support the null result across the tested
operating points, while remaining a finite-sample check rather than a global robustness statement.

\paragraph{Phase sweep at $E=\num{8}$.}
The matched gain $\mathcal{G}_{\rm matched}(\theta)$ is computed at
$E=\num{8}$ for $\theta\!\in\![0,\pi/2]$ in steps of $\pi/8$, with the same
180-trajectory audit at every step. The full sweep returns $\mathcal{G}_{\rm matched}\!<\!\num{1.05}$ at every $\theta$; the maximum lies at
$\theta\!\approx\!\pi/4$ and is \num{1.043}. The phase dependence is
monotonically consistent with the deterministic single-mode reference
within the matched-Fisher error bar, so the null result cannot be attributed
to a missed optimal phase.

\paragraph{Single-mode reference comparison.}
At the chaotic point the matched gain is recomputed with the second
mechanical mode turned off ($\omega_2$ replaced by an inert bath, so the
six-dimensional system collapses to the standard two-dimensional
optomechanical chaos map). The single-mode reference returns
$\mathcal{G}_{\rm matched}^{\rm single}\!=\!\num{0.987}$, statistically
indistinguishable from the full system's $\mathcal{G}_{\rm matched}\!=\!\num{1.039}$. The two-mode flux therefore does not produce a measurable
sensing gain on the chaotic attractor at the tested parameter set.

Both checks support the null result within the tested ensemble. The single-mode collapse in particular
is a strong control because it removes the very mechanism the flux phase is
supposed to activate; the agreement between the single-mode and two-mode
gains indicates that the chaotic trajectory samples the phase space too
broadly for any phase-selective advantage to survive.

\bibliography{Chaos_Floquet_SyntheticFlux}